\RequirePackage{fix-cm}
\documentclass[smallextended]{svjour3}       
\smartqed  
\usepackage{color}
\usepackage{graphicx}
\usepackage[dvipsnames]{xcolor}
\usepackage{soul} 
\usepackage{dcolumn}
\usepackage{amsmath, amssymb}
\usepackage{epstopdf}
\usepackage{mathtools,euscript}
\usepackage{multirow}
\usepackage{bm}
\usepackage{mathrsfs, dsfont, yfonts}
\usepackage{lipsum}
\usepackage{calc}
\usepackage{subfigure}
\usepackage{floatrow}
\usepackage{ulem}

\begin{document}

\title{Vortex elongation in outer streaming flows
}


\author{S. Amir Bahrani         \and
        Nicolas P\'erinet \and
        Maxime Costalonga \and
        Laurent Royon \and
        Philippe Brunet
}


\institute{Seyed Amir Bahrani  \at
              Universit\'e Paris Diderot, Laboratoire Mati\`ere et Syst\`emes Complexes, UMR 7057 CNRS, F-75013 Paris, France \\
              \email{amir.bahrani@imt-lille-douai.fr}             \\
             \emph{Present address: IMT Lille Douai, Univ. Lille, D\'epartement \'Energ\'etique Industrielle, F-59000 Lille, France \\} 
           \and
           Nicolas P\'erinet \at
              Departamento de F\'isica, Facultad de Ciencias F\'isicas y Matem\'aticas, Universidad de Chile, Casilla 487-3, Santiago, Chile
              \and
              Maxime Costalonga \at
              Universit\'e Paris Diderot, Laboratoire Mati\`ere et Syst\`emes Complexes, UMR 7057 CNRS, F-75013 Paris, France \\
              \emph{Present address: Department of Mechanical Engineering, Massachusetts Institute of Technology,  77 Massachusetts Avenue, Cambridge MA 02139-4307, USA\\}
              \and
              Laurent Royon \at
              Universit\'e Paris Diderot, Laboratoire Interdisciplinaire des Energies de Demain, UMR 8236 CNRS, F-75013 Paris, France
              \and
              Philippe Brunet \at
              Universit\'e Paris Diderot, Laboratoire Mati\`ere et Syst\`emes Complexes, UMR 7057 CNRS, F-75013 Paris, France \\
              \email{philippe.brunet@univ-paris-diderot.fr}   }

\date{Received: date / Accepted: date}

\maketitle

\begin{abstract}
We study the secondary time-averaged flow (streaming) generated by an oscillating cylinder immersed within a fluid, under high amplitude forcing so that inertial effects are significant. This streaming is decomposed into a viscous boundary layer flow where vorticity is created, and an outer flow of larger size. We operate under conditions of relatively low viscosity, so that  the boundary layer is smaller than the object diameter. While for low Keulegan-Carpenter (KC) number (small enough amplitude), the size of the outer flow is typically that of the object, here we show that at large enough forcing, the outer flow stretches along the direction of the vibration by up to 8 times, while the flow still keeps its axial symmetry. We quantify the elongation through PIV measurements under an unprecedented range of frequency and amplitude, so that the streaming Reynolds number reaches values much larger than unity. The absence of significant unsteady component of vorticity outside the viscous boundary layer - and the fact that the length of elongation scales well with the streaming Reynolds number - suggest that the stretching should be due to the convection of stationary vorticity by the streaming flow itself.
\end{abstract}

\section{Introduction}

\begin{figure}[H]
\centering
\includegraphics[scale=0.15]{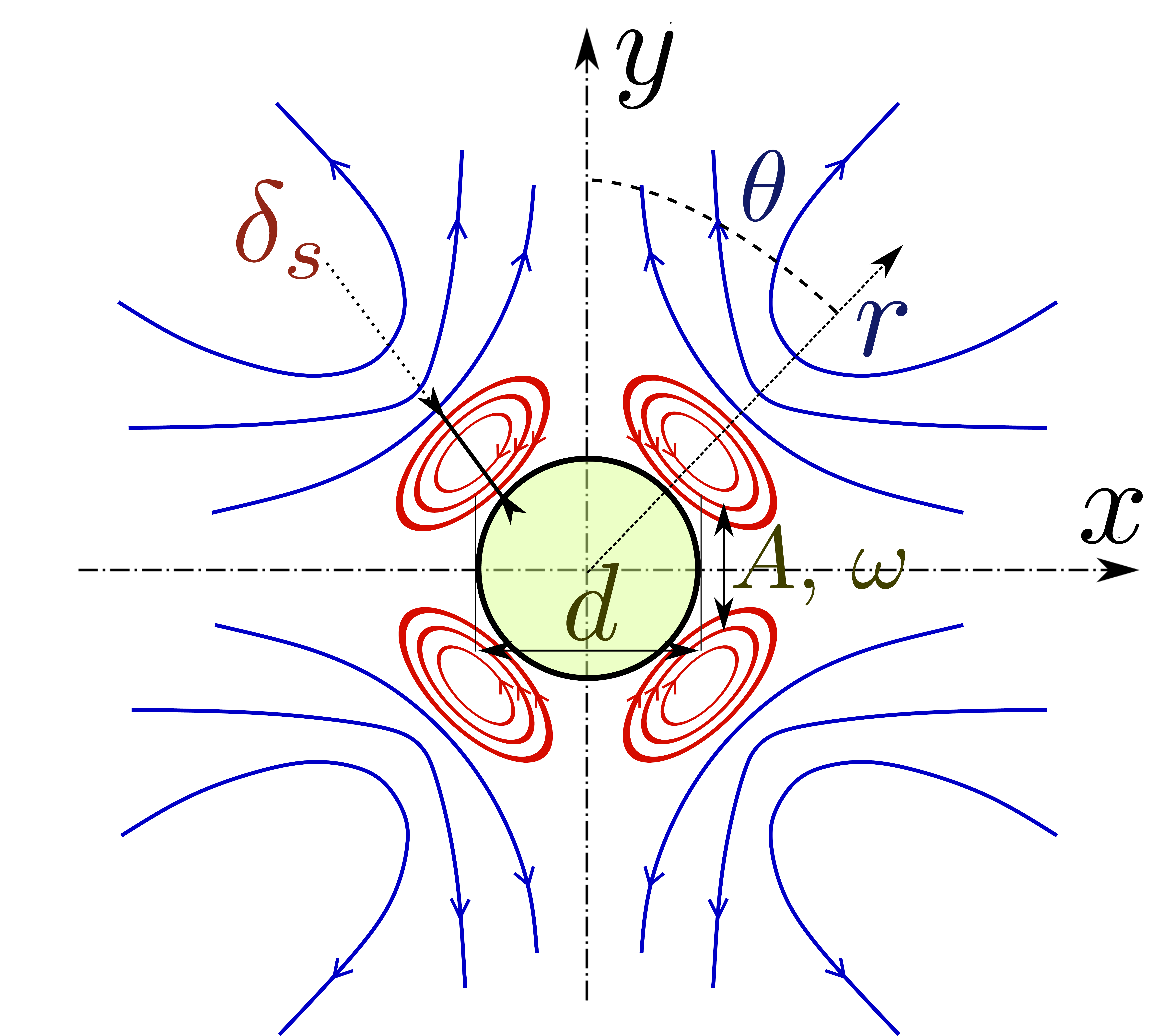}
\caption{Sketch and geometrical definitions of the secondary flow generated by the oscillations (amplitude $A$ and frequency $f$) of an immersed cylinder of diameter $d$. The inner $\vec{v}_2$ and outer $\vec{V}_2$ streaming are represented within internal and external boundary layers of respective thickness $\delta_s$ and $D$.}
\label{fig:sketch}
\end{figure}

Oscillations of bodies immersed in fluids are known to generate secondary steady flows (streaming) \cite{Riley01}, which originate from mechanisms similar to those produced by acoustic fields near solid boundaries \cite{Nyborg58}. These steady flows result from the creation of vorticity in a viscous boundary layer either around the vibrating body, or along the walls surrounding the area of sound propagation. Streaming flows have applications in fluid homogenization and mixing especially in microfluidics \cite{Suri02}, in heat transfer enhancement \cite{Loh02,Tajik13,InMeiSou11}, in particle sorting \cite{Lutz06,Devendran14} or in fluid pumping \cite{Schmid12}. This phenomenon is denoted as \textit{Rayleigh streaming}, from the pioneering study of Rayleigh on acoustically generated flows in pipes \cite{Rayleigh}, and is distinct from the \textit{Eckart streaming} originating from viscous acoustic dissipation in the bulk \cite{Eckart48}. Being localized within the boundary-layer, the first-order incompressible viscous flow generates in turn secondary streaming due to the non-linear interactions between first order viscous forces and second order inertial ones. Lighthill \cite{Lighthill78} explained the generation of streaming by volume forces resulting from spatial gradients of the Reynolds stress - either in the liquid bulk or near solid boundaries, induced by a non-zero acoustic momentum flux averaged over one period. A typical model situation, investigated in this paper, is that of an immersed cylinder oscillating perpendicular to its axis. These vibrations generate two pairs of counter-rotating vortices within the boundary layer, and by transfer of momentum and vorticity, larger eddies can be generated outside the boundary layer (\textit{outer streaming})\cite{Riley65,Stuart66,Sadhal13}. The two-dimensional (2D) flow structure is sketched in Fig.~\ref{fig:sketch}. The flow is divided into two zones, the inner boundary layer where the velocity components on $(x,y)$ are $(v_{x1} + v_{x2}, v_{y1} +  v_{y2})$ and the outer layer $(V_{x2}, V_{y2})$ (indices 1 and 2 respectively stand for the time-dependent and stationary components), whose averages over one period are $<v_{x1}> = <v_{y1}> = 0$ and $<v_{x2}>$ and $<v_{y2}>$  $\neq$ 0. Hence, the time-dependent flow should vanish away from the inner layer.

Our study aims to quantify the spatial range and strength of the outer streaming, with Particle Image Velocimetry (PIV) measurements, within a large range of frequency and amplitude; in particular within the up-to-now poorly investigated amplitude range where $A \sim d$, with $d$ the cylinder diameter. More specifically, our study focuses on the elongation (or stretching) of the outer eddies along the direction of vibration, as the streaming Reynolds number (to be defined later) is larger than a few units. Besides fundamental aspects, this effect can be of interest to induce mixing, homogenization, heat transfer and resuspension of particles far from the vibrating object.

The steady flow around a vibrating cylinder was theoretically investigated in various studies \cite{Riley65,Stuart66,Holtsmark54,Milton_Andres53,Davidson_Riley72} and summarized in the monographs of Batchelor \cite{Batchelor} and Schlichting \cite{Schlichting} - who additionally investigated the first-order flow in the oscillating boundary layer. By taking $\vec{V}(t) = V_0 ~ \sin (\omega t) ~\vec{e_y}$ (where $V_0 = A \omega$  and the phase $\phi = \omega t$), as the time-dependent velocity of the vibrating cylinder, $A$ and $\omega$ being the amplitude and angular frequency, these studies predicted a time-averaged secondary flow $(v_{x2},v_{y2})$ within a boundary layer of thickness $\delta_s$ around the object, depending on the space variable $y$, and expressed in Cartesian coordinates as:

\begin{eqnarray}
<v_{x2}>  & = & \frac{3}{4 \omega} \frac{d}{dy} (V_{0c} \frac{d V_{0c}}{d y}) \left(y - \frac{13}{6} \delta \right) \\
<v_{y2}> & = & - \frac{3}{4 \omega} V_{0c} \frac{d V_{0c}}{d y}
\label{eq:streaming_cyl1}
\end{eqnarray}

\noindent where $\delta = \sqrt{\frac{2\nu}{\omega}}$ stands for the thickness of the oscillating boundary layer, which can be slightly different from its steady counterpart $\delta_s$ \cite{Holtsmark54,Stuart66}. Here, $V_{0c}$ is taken as the component of the vibration velocity that is normal to the surface of the cylinder, i.e. $V_{0c}$ is maximal at the vibration axis ($\theta$=0 and $\pi$) and $V_{0c}$ = 0 at the nodes of the vibration ($\theta = \pm \frac{\pi}{2}$). For any point along the cylinder $(r_c,\theta_c)$ or $(x_c, y_c)$, $V_{0c} = A \omega \cos (\theta_c) = A \omega \frac{y_c}{\left(x_c^2+y_c^2 \right)^{1/2}}$. Hence, it yields: $\frac{d V_{0c}}{d y_c} = \frac{A \omega}{r_c}$. Incidentally, it is remarkable that the streaming flow is related to the non-zero local curvature of the vibrating object. This is reminiscent of the streaming induced by acoustic fields in microchannels either around cylindrical or squared posts \cite{Lieu12}, or near sharp edges \cite{Nama14,Ovchinnikov14,Zhang19} which concentrate the streaming force.

When $\frac{\delta}{d}$ is small enough, the time-averaged velocity field outside the boundary layer $(V_{x2},V_{y2})$ can be deduced from the inner flow by taking the inner streaming velocity $(v_{x2},v_{y2})$ along the periphery of the boundary layer as a slip velocity boundary condition \cite{Nyborg58,Sadhal13}. Eq.~(\ref{eq:streaming_cyl1}) says that the vertical component of the inner flow $v_{y2}$ is maximal at $x=0$. The outer flow is such that the fluid is pushed away from the vibration antinodes and flows toward vibration nodes, see Fig.~\ref{fig:sketch}. Stuart \cite{Stuart66} analyzed theoretically the flow both inside and outside the boundary layer, and showed that the size of outer vortices decreased with the forcing amplitude. \textcolor{ blue}{In the related case of vibrating spheres, Riley \cite{Riley1966} and Amin and Riley \cite{Amin_Riley1990} investigated the secondary streaming flow and provided analytical expressions for low to moderate streaming Reynolds numbers. In \cite{Amin_Riley1990} qualitative experiments were conducted, suggesting that the classical streaming flow progressively turns to jet-like flow shooting from the axis of vibration.}

The analyses of secondary flows generally introduce dimensionless parameters: the Keulegan-Carpenter ($KC$) number compares the amplitude $A$ with the size of the object $KC = \frac{2 \pi A}{d}$ (also appearing in the acoustic literature as: $\epsilon = \frac{A}{d} = \frac{KC}{2 \pi}$), the Stokes number $\beta = \frac{d^2 \omega}{2 \pi \nu} = \left(\frac{d}{\delta} \right)^2$ is the square of the ratio between the cylinder diameter and the boundary layer thickness. The KC number is the inverse of the Strouhal (St) number: $\text{St} = (\text{KC})^{-1}$, which is most often found in von-Karman flows studies. The Stokes number $\beta$ is also related to the Womersley (M) number ($\beta = \frac{1}{2 \pi} M^2$) \cite{Gondret_Petit}. The Reynolds number can be defined with respect to the object vibration: $Re_1 = \frac{A\omega d}{\nu} = \frac{V_0 d}{\nu}$, or with respect to the secondary flow: $Re_s = \frac{U_{\text{str}} d}{\nu}$, with $U_{\text{str}}$ the characteristic velocity of the streaming flow.

The streaming flow was experimentally investigated in various studies with immersed vibrating cylinders or spheres \cite{Riley65,Stuart66,Holtsmark54,Davidson_Riley72,Gondret_Petit,Bertelsen73,Tatsuno77,Tatsuno90,Kotas07,Kotas08}, or for a static cylinder immersed in an oscillating flow \cite{Holtsmark54}. Bertelsen \textit{et al.} \cite{Bertelsen73} investigated the inner and outer streaming flow profiles, showing that the outer vortices grow in size with the size of the external container and evidencing the influence of outer boundary conditions. Tatsuno \cite{Tatsuno77} showed that in a low Reynolds number ($Re_1$) situation (i.e. $\delta \gg d$), the size of inner vortices decreased with amplitude, and that the outer flow was significantly influenced by both the amplitude and the size of the container. Tatsuno and Bearman \cite{Tatsuno90} studied flow regimes at higher amplitude ($A > d$) and inviscid flows ($\delta \ll d$), and identified a host of time-dependent and asymmetrical regimes (e.g. vortex shedding and transition to von Karmann vortices), but without straightforwardly pointing out the phenomenon of vortex elongation along the vibration axis. Still, to the best of our knowledge, none of these studies quantitatively investigated the size of the outer flow in the range where $Re_s$ is larger than a few units. More recent studies of secondary flows generated by vibrating spheres were conducted in a larger range of Re$_s$ number \cite{Blackburn2002,Klotsa2007,Klotsa2009,Otto2008}. In the paper by Blackburn \cite{Blackburn2002}, the mass transport and drag of oscillating sphere was investigated from Re$_s$=1 to Re$_s$=100, and the narrowing of eddies was noticed at large Re$_s$. In Voth \textit{et al.} \cite{Voth2002} and Klotsa \textit{et al.} \cite{Klotsa2007,Klotsa2009}, the formation of clusters of non-brownian particles by streaming flows was investigated experimentally and numerically, revealing the remarkable periodicity of the patterns of particles \cite{Klotsa2007}, and investigating the underlying mechanisms through the study of transient behavior \cite{Klotsa2009}.

\begin{table}[ht]
		\begin{tabular}{ccc}
			Symbol & Physical quantity & Units \\
			\hline
			$\rho$ & Liquid density & kg.m$^{-3}$ \\
			$\eta$ & Liquid dynamic viscosity & Pa.s \\
			$\nu$ & Liquid kinematic viscosity & m$^2$.s$^{-1}$ \\
			$A$ & Cylinder vibration amplitude & m \\
			$f, \omega$& Cylinder vibration frequency and angular frequency & Hz, rad.$s^{-1}$ \\
			$d$ & Cylinder diameter & m\\
			$\epsilon$ & Dimensionless amplitude & - \\
			$ V_0$ & Vibration velocity & m/s \\
			$V_{0c}$  & Vibration velocity component normal to the vibration axis & m/s  \\
			$\delta$ & Thickness of unsteady inner BL & m \\
			$\delta_s$ & Thickness of steady inner BL & m \\
			$D$ & Thickness of steady outer BL & m \\
			$\vec{v_1}$ & Unsteady velocity & m/s \\
			$\vec{v_2}$ & Steady velocity & m/s \\
			$\Omega_1$ & Unsteady vorticity & s$^{-1}$ \\
			$\Omega_2$ & Steady vorticity & s$^{-1}$ \\
			$p_1$ & Unsteady pressure & Pa \\
			$p_2$ & Steady pressure& Pa \\
			$U_{str}$ & Typical streaming velocity & m.s$^{-1}$ \\
			$v_{max}$ & Maximal streaming velocity & m.s$^{-1}$ \\
			$v_{y,max}$ & Maximal vertical streaming velocity & m.s$^{-1}$ \\
			$y_{max}$ & Vertical position of the maximal streaming velocity & m.s$^{-1}$ \\
			$r_a$ & Radial position of the center of the flow streamlines & m \\
			$\theta_a$ & Angular position of the center of the flow streamlines & - \\
			$r_m$ & Radial position of the maximal vorticity & m \\
			$\theta_m$ & Angular position of the maximal vorticity & rad \\
			$l_a$ & Typical length range for the streaming flow & m \\
			$\theta$ & Angular direction for the maximal length range & rad \\
			KC & Keulegan-Carpenter number & - \\
			$\beta$ & Stokes number & - \\
			St & Strouhal number & - \\
			M & Womersley number & - \\
			Re & Reynolds number & - \\
			Re$_s$ & Streaming Reynolds number & - \\
			
		\end{tabular}
			\caption{\label{table1} Definitions and symbols of the different parameters and quantities.}
\end{table}

The elongation we report in this study, cannot be explained within the usual frame of low-amplitude vibrations. Indeed, the classical approach predicts that the outer vortices naturally scale like the size of the vibrating object. Our experiments show that the size of flow structures can increase by a factor of up to 8 along the vibration axis. Therefore, we carried out an experimental analysis of the flow in order to quantify and understand better the mechanism of the stretching. \\
The paper is organised as follows: in section II the experimental setup is presented, in section III we first review expected scaling laws, and show a comparison of experiments with theoretical predictions in the regime of weak forcing, then extended to larger amplitude forcing. Then we propose possible interpretation in section IV and we conclude on the main outcomes of our results.

\section{EXPERIMENTAL SETUP}

\begin{figure}[H]
\centering
\subfigure[]{\includegraphics[scale=0.26]{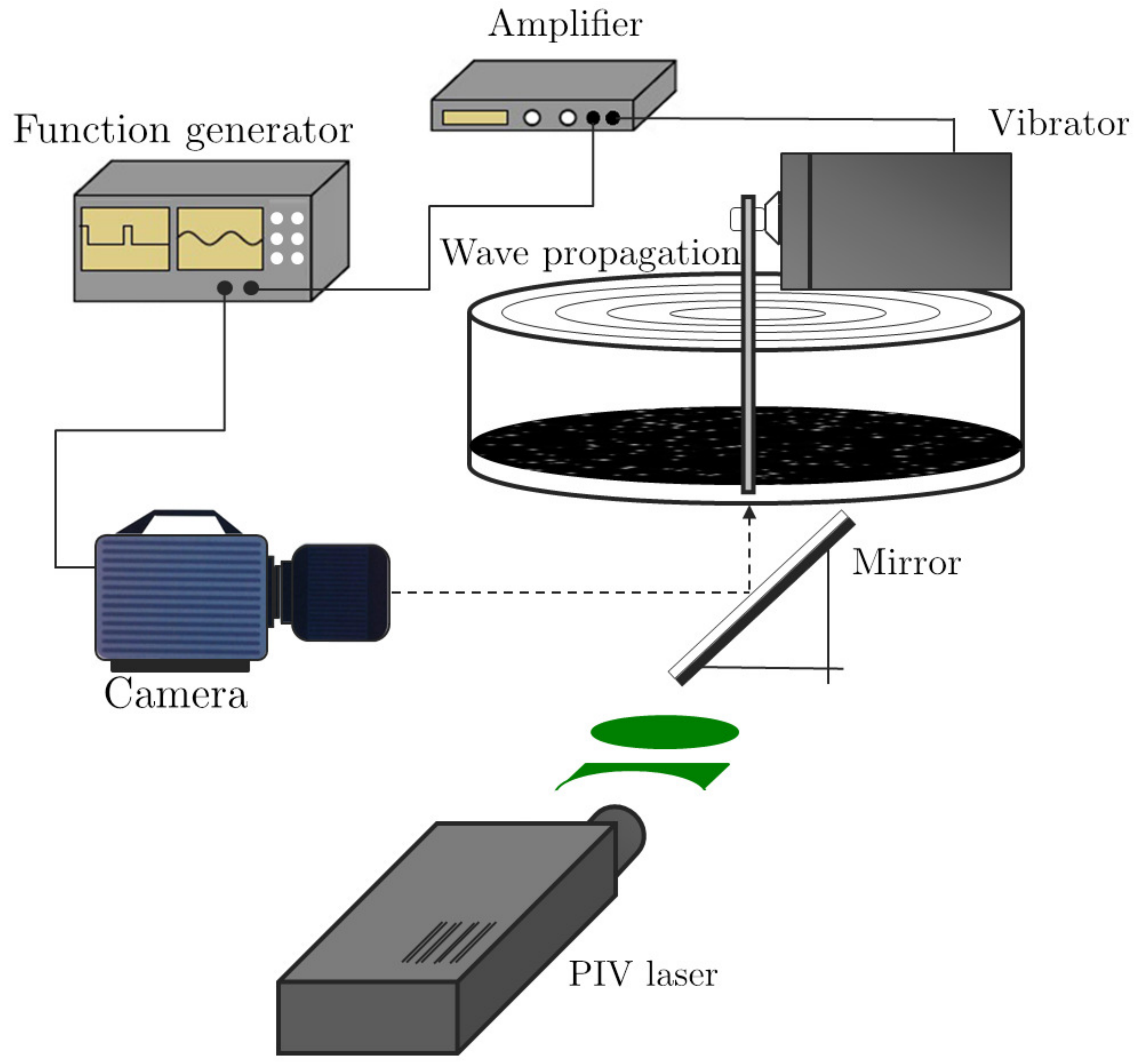}}
\subfigure[]{\includegraphics[width=0.42\textwidth]{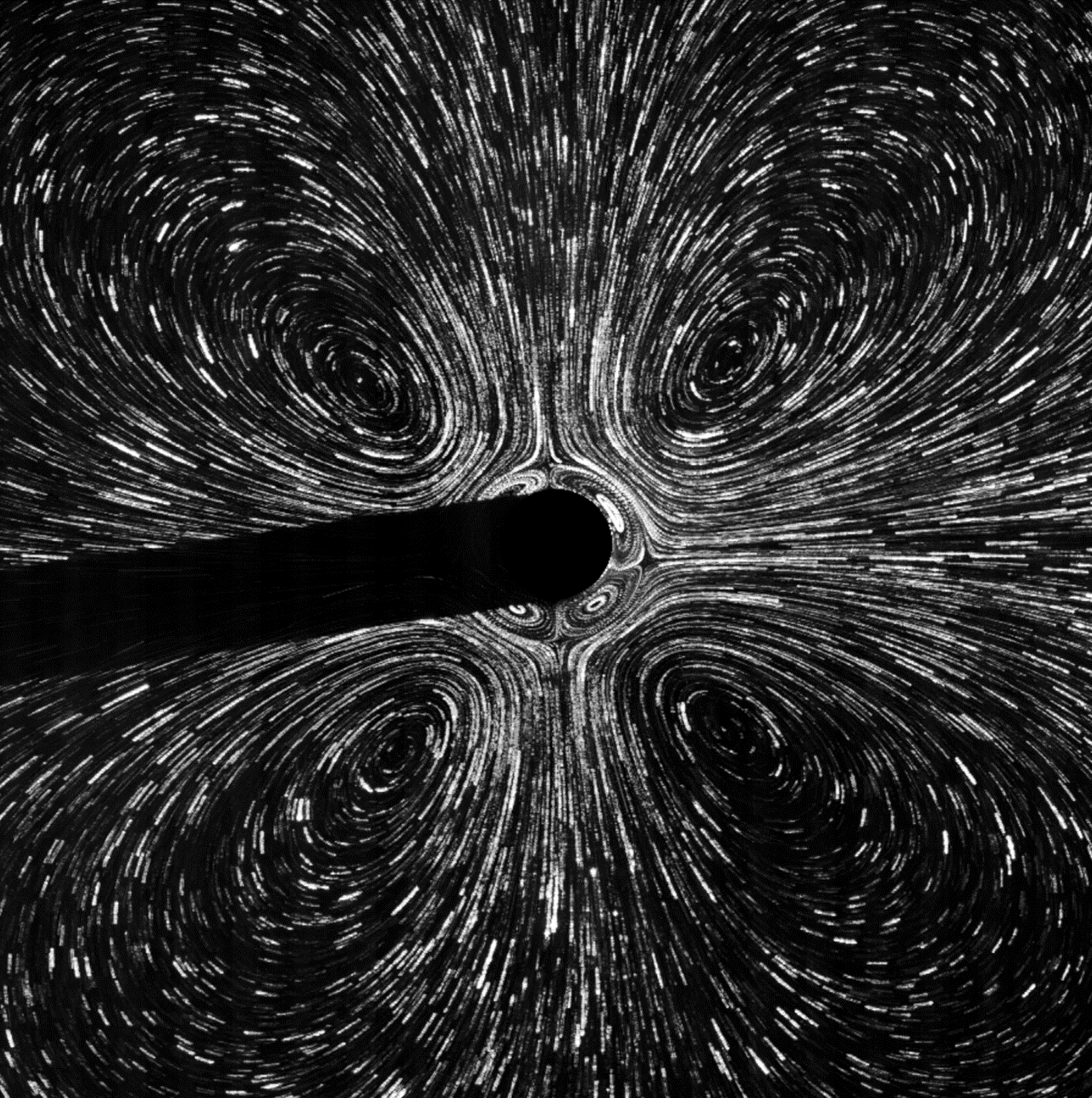}}
\caption{(a) Sketch of the experimental system, see text for details. (b) The structure of the secondary flow is revealed by trajectories of particles using long-time exposure (several tens of periods). The black region in the left, is the shadow of cylindrical beam (diameter $d$= 5 mm). Close to the cylinder, one sees the four inner vortices.}
\label{fig:setup}
\end{figure}

The experimental system (Fig. \ref{fig:setup}-(a)) consists of a aluminium cylinder beam, denoted as the vibrating object, of diameter $d$ = 5 mm, mounted on an electromechanical vibrator (minishaker type 4810, Br\"{u}el \& Kj\ae r, N\ae rum, Denmark). This vibrating object is immersed in a fluid and shaken perpendicularly to its axis. The cylindrical container has a diameter $l$ = 13 cm and the height of the fluid is $L$ = 8 cm.
The fluid is silicon oil (Polydimethylsiloxane, Sigma Aldrich) of dynamic viscosity $\eta$ = 9.3$\times$10$^{-3}$ Pa.s, kinematic viscosity $\nu$ = 10 cSt, density $\rho$ = 930 kg/m$^3$ and surface tension $\sigma$ = 0.0205 N/m at 25$^{\circ}$C.
A periodic sinusoidal voltage, generated by a function generator (DG4062, RIGOL Technology Inc., USA), is transmitted to the vibrator via a power amplifier (type 2718, Br\"{u}el \& Kj\ae r, N\ae rum, Denmark). It allows us to finely tune the amplitude $A$ (corresponding to a net displacement of $2A$ over one period) and frequency $f = \omega / 2\pi$ of the vibrations, which are the two main control parameters here. Table 1 lists the physical quantities and parameters used in this study. The amplitude was determined both by direct visualisation and by an accelerometer (PCB Piezotronics mode 256HX) fixed to the cylinder which, via the Fourier analysis on an oscilloscope (Rigol DS1102E) also allowed to check that the forcing remains sinusoidal.

The evolution of the flow structure is determined by PIV. 
The fluid is therefore seeded with silver-coated hollow-glass spheres of 10 $\mu$m diameter (Dantec Dynamics) of density $\rho_s$ = 1060 kg/m$^3$. An ($r$--$\theta$) plane is illuminated by a Nd:Yag laser sheet ($\lambda$=532 nm, thickness 1 mm). The flow pattern is recorded with a high-speed camera synchronized with the function generator (Phantom V7, Vision Research Inc., USA) - by means of a mirror oriented at an angle of 45$^{\circ}$ with the horizontal. We choose the recording frame rate to be 10 to 20 times the frequency of the vibrations, allowing us to  capture both the first-order oscillatory flow and the streaming flow. In the same way as in a previous study \cite{Costalonga15}, a PIV treatment macro is applied to series of two images with a time interval of one period between them, hence taken at the same phase. We add several of these successive iso-phase images to obtain the traces of individual particles, which is equivalent to long-time exposure shots. A typical result is depicted in Fig.~\ref{fig:setup}-(b): it is pictured out from long-time exposure recording (typically 20 to 100 periods). In this image, and in all the following images, the vibration is always oriented along the vertical axis.

Since vibrations can induce waves on the liquid surface, we carefully checked that these waves did not disturb the bulk flow. We checked that the measured flow is independent of the vertical position of the laser sheet - when it is far enough from the container base and the free-surface, in practice at a distance of roughly 2 cm from the bottom surface. 

\section{EXPERIMENTAL RESULTS}

\subsection{Flow analysis}

Tatsuno and Bearman \cite{Tatsuno90} and Elston \textit{et al.} \cite{Elston06} investigated the different regimes generated by a vibrating cylinder within a large range in $KC$ and $\beta$. They showed that for low enough $KC$ and $\beta$ (the threshold on each parameter depends on the other one), there is no vortex shedding, no flow separation nor tridimensional flow, i.e.:

- the flow is symmetric with respect to the vibration axis and invariant along the direction of the cylinder axis. 

- the averaged first-order flow is zero, so that the time-averaged flow is only streaming. 

The corresponding flow was denoted as $\mathcal{A}$* and $\mathcal{A}$ regimes, and we keep this denomination for the sake of consistence. The transition between these two regimes were described in \cite{Tatsuno90} as when \textit{"the flow is composed of secondary streaming and flows due to vortices resulting from separation"}. They associated this transition to the appearance of vortex shedding, which is visible along the vibration axis (see Fig. 3 of \cite{Tatsuno90}).

Thus, we choose our experimental range in the $\mathcal{A}$* and $\mathcal{A}$ regimes, so that the outer flow consists in four vortices, whose maximal intensity is located at some distance $r_m$ and angular position $\theta_m$ from the centre of the cylinder. In practice, the amplitude is kept smaller than half of the object size, i.e. $A \le $ 2.5 mm, within the frequency range between 5 and 60 Hz. Let us mention that due to the limitation of the shaker and amplifier power, the maximal value of 2.5 mm could be prescribed only for moderate frequencies (roughly, for $f \le $ 25 Hz).

\begin{figure}[H]
\centering
\includegraphics[width=1\textwidth]{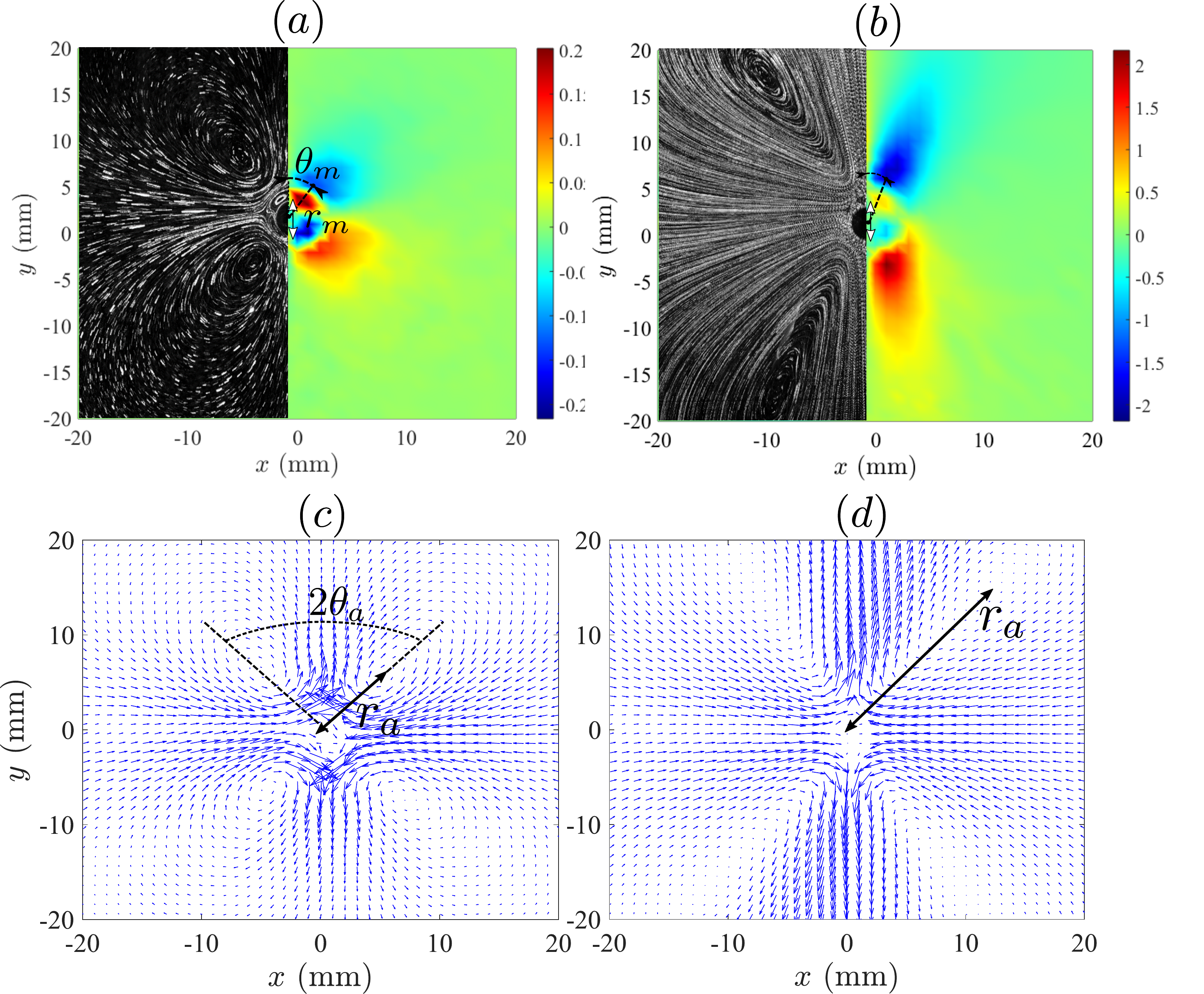}
\caption{(a,b) Long-time exposure visualisation of particle trajectories for the steady streaming flow (left) and their corresponding vorticity maps (right) and (c,d) velocity fields of the streaming flow, for $f$=25Hz and $A$= 0.5 mm (a,c) and $f$=25Hz ($V_0$ =0.078 m/s, $\beta$ = 62.5, KC = 0.628, Re$_s$ = 0.625)  and $A$= 1.8 mm (b,d) ($V_0$ =0.28 m/s, $\beta$ = 62.5 , KC = 2.262, Re$_s$ = 8.1). The straight line connects the center of the cylinder to the center of the flow streamlines $(x_a,y_a)$.}
\label{fig:piv_vort}
\end{figure}

In the $\mathcal{A}$* and $\mathcal{A}$ regimes, the vortices of the secondary flow are symmetric over the $y$ axis, and the two pairs are symmetric over the $x$ axis. This is also true for the center of the recirculation zones, as revealed by the long-time exposure recording, see Fig.~\ref{fig:setup}-(b). However, it does not imply that the angles corresponding to maximal and minimal vorticity be equal to 45$^{\circ}$. Although for weak enough forcing $\theta \simeq 45^{\circ}$ (see Fig.~\ref{fig:setup}-(b) and Fig.~\ref{fig:piv_vort}-(a)), this is no longer true at higher forcing, as shown in Fig.~\ref{fig:piv_vort}-(b). The flow structure is no longer symmetric over the $x=y$ and $x=-y$ axes. The distance ($r_a$) between the centre of the cylinder and that of the flow streamlines (i.e. the location of the eye of the vortices) increase with the quantity $A^2 f$. These recirculation zones are pushed away from the ($x$=0) axis, and this seems to be the difference between $\mathcal{A}$* and $\mathcal{A}$ regimes underlined in Tatsuno and Bearman's study \cite{Tatsuno90}. From Figs.~\ref{fig:piv_vort}-(a-d), it is clear that $r_a$ and $r_m$ are different, and so are $\theta_a$ and $\theta_m$.

Simple scaling arguments lead to fair predictions for the velocity of the secondary flow, at moderate forcing and for unconfined geometry. Let us plug the unsteady $\vec{v_1}$ and steady $\vec{v_2}$ velocity fields. We also operate the same decomposition for the vorticity $\Omega = \Omega_1 + \Omega_2$. Considering $ \Vert \vec{v_2} \Vert \ll \Vert \vec{v_1} \Vert$ as a small perturbation of the primary flow, and taking the time-average of the incompressible Navier-Stokes equation, it yields: 

\begin{equation}
< \rho (\vec{v_1} . \vec{\nabla})~\vec{v_1}> = - \vec{\nabla} p_2  + \rho \nu~\Delta \vec{v_2}
\label{eq:ns_o2}
\end{equation}

The nonlinear term involving $\vec{v_1}$ can be accounted for an effective volume force, which is non-zero only in the boundary layer. Seeking for a scaling law, one sets $\Vert \vec{v_1} \Vert \sim A \omega$ and the diameter $d$ as being the length scale for the variations of $\vec{v_1}$. The thickness $\delta$ is chosen as the scale to the gradient of $\vec{v_2}$. Hence, considering only the vertical component of the time-averaged flow ($v_2$), leads to $ \nu \frac{v_2}{\delta^2} \sim \frac{(A \omega)^2}{d} $, which yields the following scaling for $v_2$:

\begin{equation}
v_2 \sim \frac{A^2\omega}{d}
\label{eq:scaling_v2}
\end{equation}

\noindent that can also be retrieved from eq.~(\ref{eq:streaming_cyl1}). Let us remark that taking $U_\text{str} =  v_2$ in the expression of $Re_s$ yields: $Re_s = \left(\frac{A}{\delta} \right)^2$, which becomes larger than one when the amplitude overcomes the thickness of the  boundary layer. Pursuing this scaling approach for vorticity, from the evaluation of the circulation of $v_2$ around a vortex, one can write: $v_2 d \simeq \Omega d^2$, which comes from the Kelvin-Stokes theorem stating the equality between the circulation of $v_2$ and the flux of vorticity. This assumes the vortex size to be equal to $d\times d$, which is valid only for low enough forcing, as we will see in more details later. Therefore, a characteristic value for the steady vorticity $\Omega_2$ reads:

\begin{equation}
\Omega_2 \sim \frac{A^2\omega}{d^2}
\label{eq:scaling_omega2}
\end{equation}

Consistently with eq.~(\ref{eq:streaming_cyl1}), the streaming flow is assumed to be independent of viscosity. This independence holds as long as the thickness of the inner boundary layer $\delta_s$ remains small compared to the other dimensions of the problem, like the cylinder diameter $d$ or the container size \cite{Costalonga15}.

Stuart's analysis \cite{Stuart66} predicted that when $Re_s$ is larger than a few units, a second external boundary layer would exist, due to convection of vorticity from the inner boundary layer, and that the size of these outer vortices $D = d \left(\frac{\omega \nu}{V_0^2}\right)^{1/2} = d \left(\frac{\nu}{A^2 \omega}\right)^{1/2} = d (Re_s)^{-1/2}$ would decrease with $A$, within the range $KC \ll 1$. 

\begin{figure}[H]
\centering
\includegraphics[width=1\textwidth]{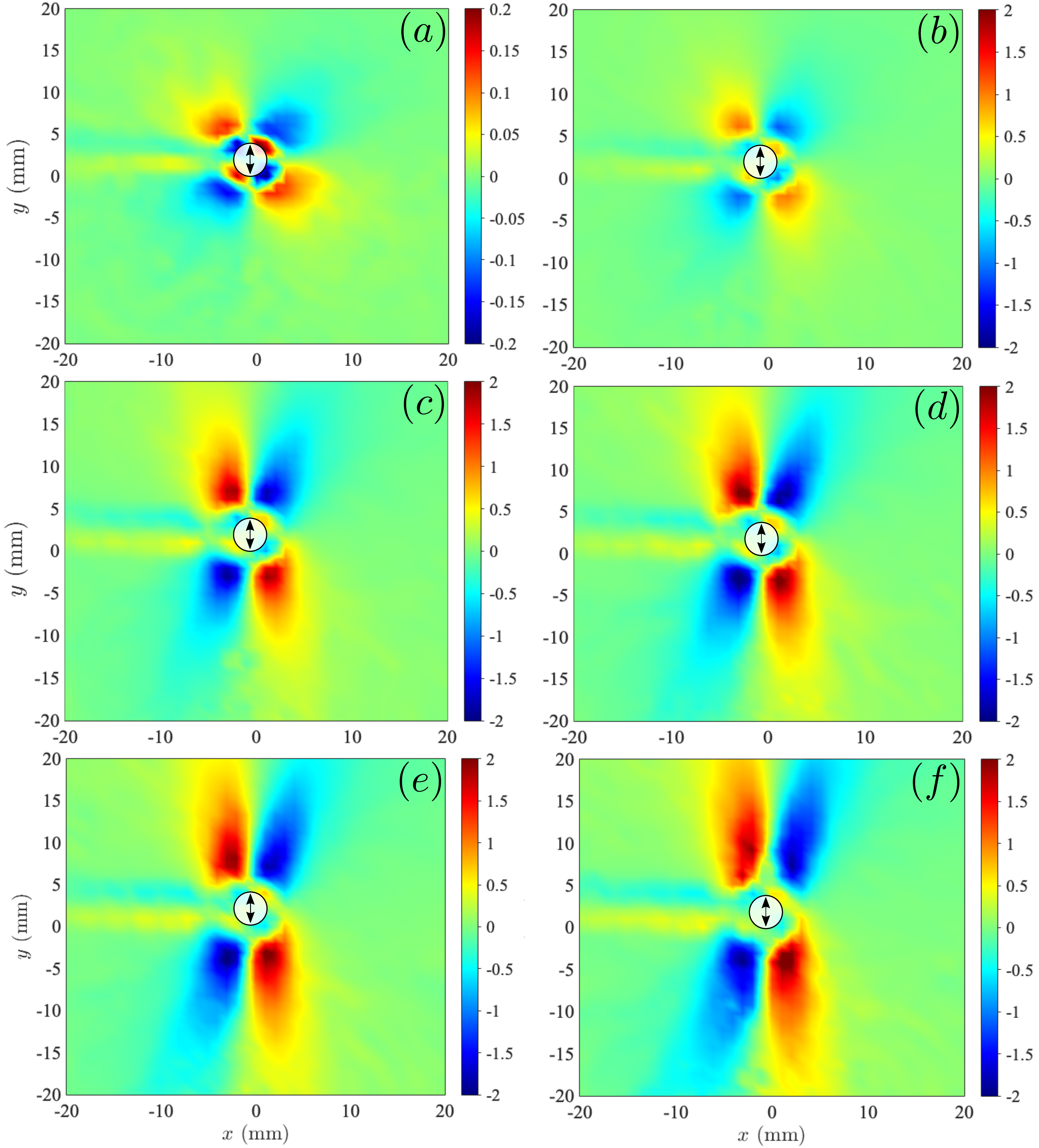}
\caption{Vorticity maps at $f$=25 Hz, for different amplitudes $A$ of increasing values from (a) to (f). (a) $A$= 0.46 mm, (b) 1.19 mm, (c) 1.46 mm, (d) 1.61 mm, (e) 1.80 mm and (f) 1.88 mm. Let us note that the colormap scale of figure (a) is ten times smaller than those of figures (b-f).}
\label{fig:evol_vort_25Hz}
\end{figure}

Figure \ref{fig:evol_vort_25Hz} shows vorticity maps for different increasing values of amplitudes $A$, at the same frequency of 25 Hz. It clearly illustrates that in the lower range of $A$, an increase of $A$ leads to an intensification of vorticity without significant elongation, while in the upper range of $A$ this increase leads to the vortex elongation while the vorticity saturates. Let us notice that the shadow of the cylinder is visible by artefacts colors in the left side of the cylinder. From these series of vorticity maps, we can 
infer that vortex elongation occurs beyond a threshold in amplitude, after a first phase of vorticity intensification with A increasing within its lower range.
 
\subsection{Instantaneous flow over a period of oscillation}

We aim to relate the behavior of the streaming eddies and their stretching at high enough $Re_s$ with that of the instantaneous unsteady flow. The time-periodic flow is measured with PIV technique similar to that employed for the streaming flow. However, it is determined from visualisations with stronger magnification than that used for the streaming flow, to reveal the flow around the cylinder. Also, the velocity fields taken at different fixed phases are determined from an average over 10 consecutive periods, in order to smooth or suppress some unphysical noise, especially close to the cylinder. 

\begin{figure}[H]
\centering
\includegraphics[scale=0.27]{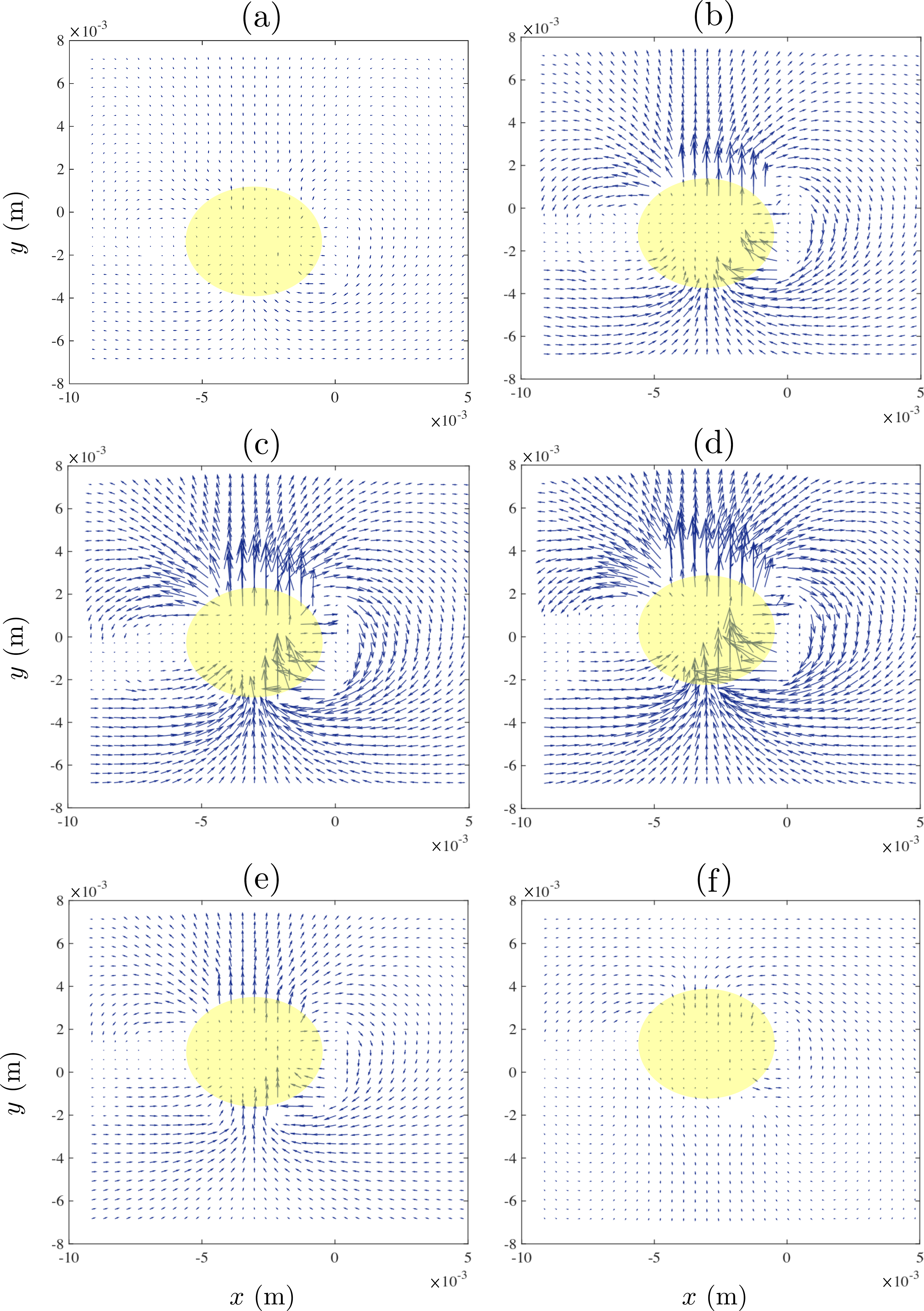}
\caption{Successive unsteady velocity fields around the oscillating cylinder ($A$=1.31 mm, $f$ = 40 Hz, giving $A \omega$ = 0.33 m/s), represented in light yellow. Only the first half period is shown. Ticks on axes denote the location in meters. From (a) to (f), the successive times are 0 ms, 2.5 ms, 5 ms, 7.5 ms, 10 ms, 12.5 ms (corresponding to phases $\phi$=0, $\pi/5$, $2\pi/5$, $3\pi/5$, $4\pi/5$, $\pi$), the origin of phase and time being taken at the lowest vertical location of the cylinder (when $V(t)$ is null).}
\label{fig:unst_flow}
\end{figure}

Figure \ref{fig:unst_flow} shows an example of successive instantaneous velocity fields around the oscillating cylinder. These fields correspond to a typical situation when the stretching of vortices is observed, with $A$ = 1.31 mm and $f$=40 Hz ($Re_s$=6.86). The sequence starts from the (a) subfigure that corresponds to the lowest position of the cylinder ($\phi$=0), and ends to the (f) one that corresponds to the highest position of the cylinder ($\phi = \pi$). The flow is more intense around the phases $\phi = 2\pi/5$ and $\phi = 3\pi/5$, indeed shortly before and after the cylinder maximal velocity at $\phi = \pi/2$, corresponding to the cylinder median position. This should corresponds to the phases of maximal vorticity. Conversely, the unsteady flow is very weak when the cylinder lies at its lowest and highest positions along the $y$ axis, which should correspond to the phase of minimal vorticity. Obviously, the inner unsteady flow is in phase with the cylinder oscillations. The PIV sequences are taken at 20 frames per period. For instance, in the example of Fig.~\ref{fig:unst_flow} of oscillations at $f$=40 Hz, we opt for 800 fps. This constitutes a fair compromise between the time-sampling accuracy and the ability to measure a consistent velocity in the region of space of a few cm around the cylinder. The maximal value of velocity reaches 0.3 m/s, rather close to the maximal velocity of the cylinder in the laboratory frame, here $A \omega$ = 0.33 m/s. 

\begin{figure}[H]
\centering
\includegraphics[width=1\textwidth]{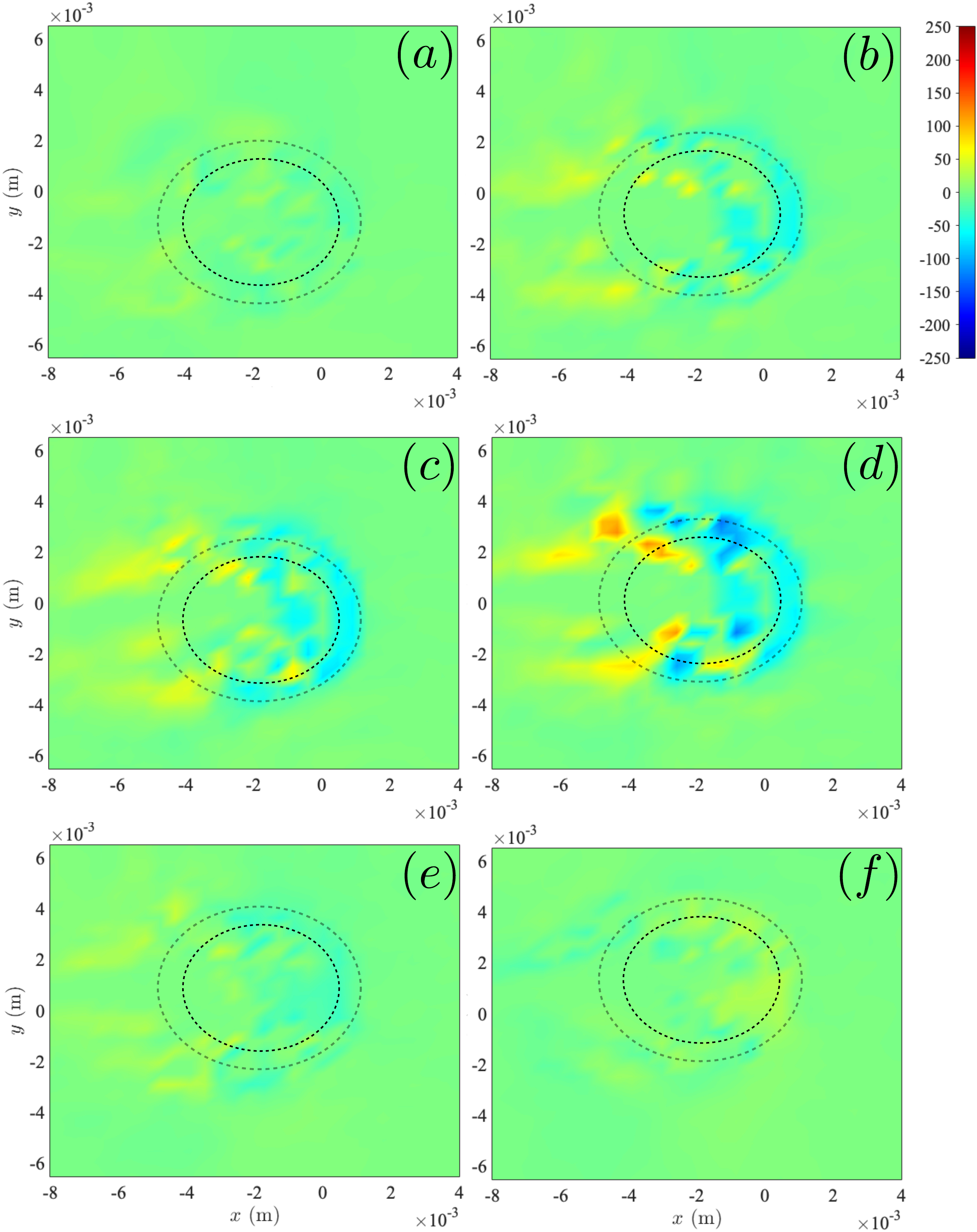}
\caption{Unsteady vorticity fields around an oscillating cylinder, represented here as a dashed ellipse. The outer dashed ellipse represents the limits of the viscous boundary layer. Conditions are those of Fig.~\ref{fig:unst_flow}. The colour bar scale is the same for all plots. Successive times and phases are the same as in Fig.~\ref{fig:unst_flow}.}
\label{fig:unst_vort}
\end{figure}

We then determine the unsteady vorticity $\Omega_1$ at the corresponding phases, see Figure \ref{fig:unst_vort}. As in previous maps, colors code for the intensity of vorticity in the $z$ direction normal to the plane. For sake of clarity, we took the same color bar scale for all pictures. From these maps, we can clearly state that the unsteady vorticity is mainly generated within the viscous boundary layer. This is expected from general considerations on unsteady boundary layers \cite{Schlichting,Riley65,Stuart66,Sadhal13}, but was worth being confirmed in the situation of large amplitude forcing, where $A$ can be several times larger than $\delta$. Then, although significant unsteady velocity is shown out of the boundary layer, this unsteady component is irrotational.

As supplementary information, we provide movies edited from these two fields, over one entire period of oscillations, taken at the 20 different phases ($\phi$ = 0 to 19$\pi$/10, with increment $\delta \phi$ = $\pi$/10).

\subsection{Streaming flow: quantitative analysis} 

Typical 2D-maps of the streaming velocity ($v_{2x},v_{2y}$) and vorticity ($\Omega_{2z}$) fields are shown in Figs.~\ref{fig:piv_vort}-(a-d) for a given frequency $f$ = 25 Hz and two values of amplitude $A$. We first extract quantitative values from PIV measurements. 

\begin{figure}[H]
\centering
\subfigure[]{\includegraphics[scale=0.32]{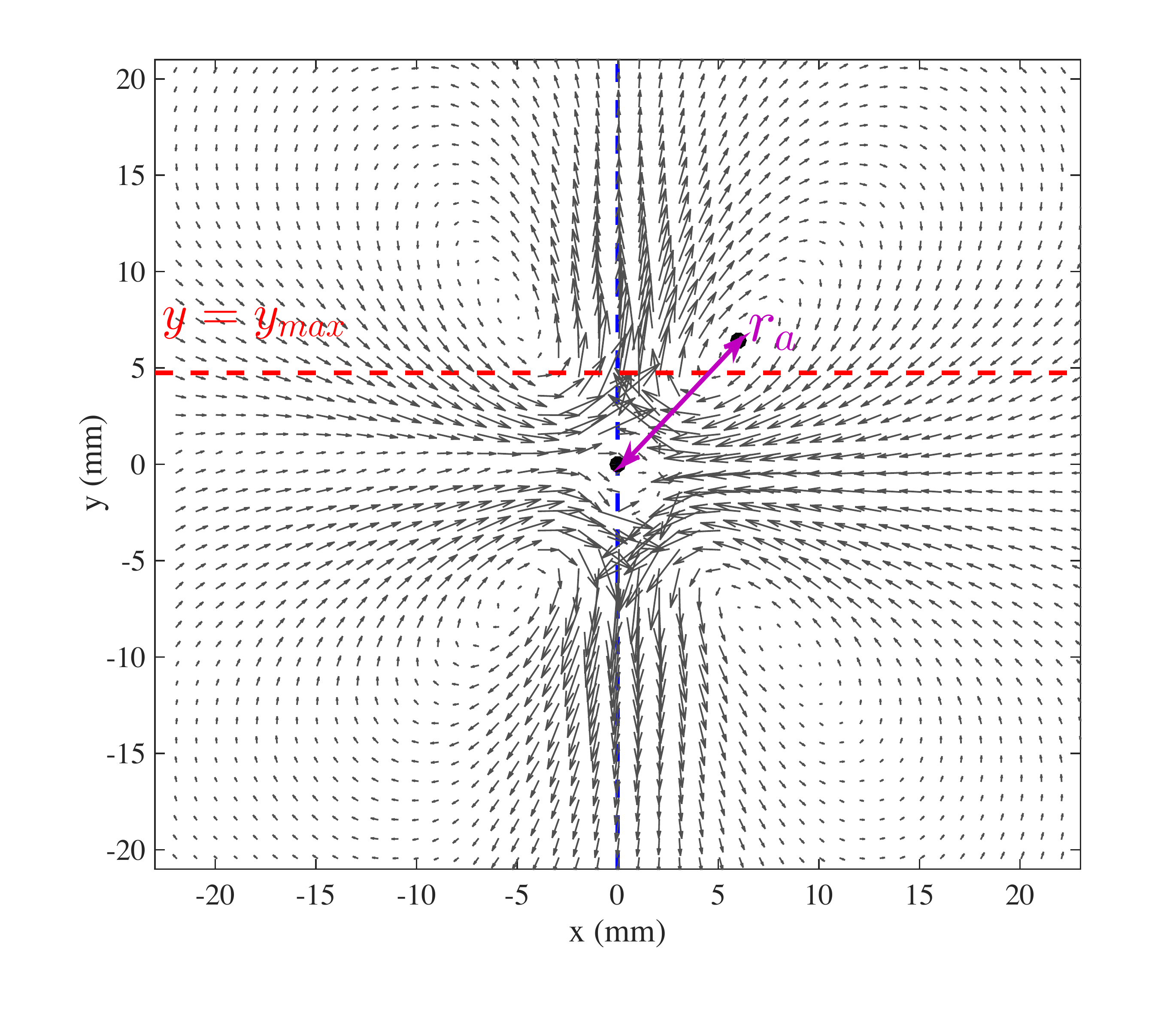}}
\subfigure[]{\includegraphics[scale=0.42]{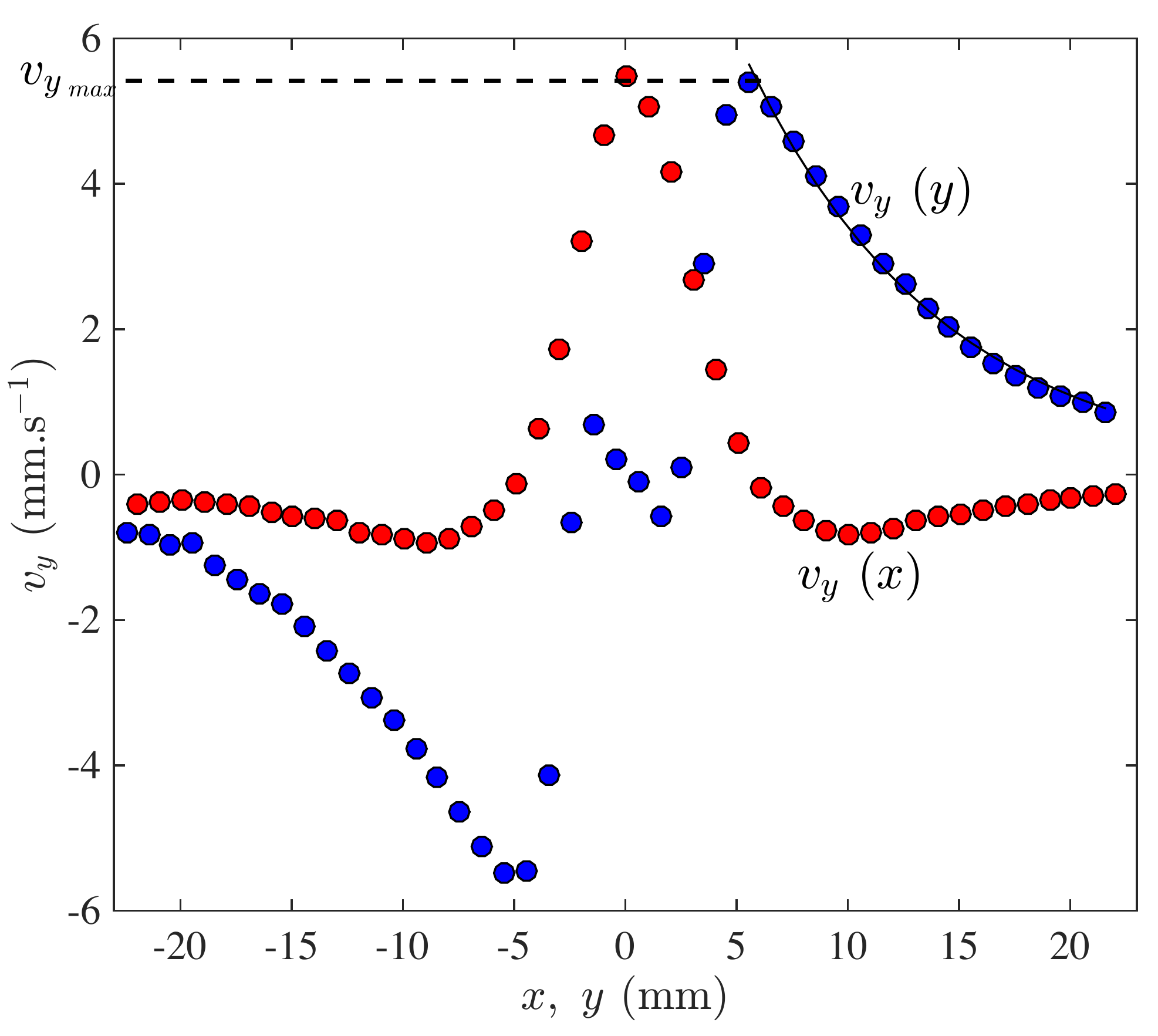}}
\caption{(a) Sketch of the quantitative extraction of velocity profiles along x and y axes, from 2D PIV measurements. (b) Example of resulting profiles for the vertical streaming velocity $v_y$ versus $y$ along the axis $x=0$ (blue disks), which enables to determine $v_{y, \text{max}}$ and versus $x$ along the axis $y=y_{\text{max}}$ (red disks). The solid black line is a fitting curve allowing to determine the characteristic length of decay of the streaming flow $l_a$ from eq.~(\ref{eq:fit_Vy}).}
\label{fitting_curve}
\end{figure}

The maximum velocity along the vibration axis is extracted from the velocity field, as sketched in Fig.~\ref{fitting_curve}-(a). We obtain the profile of $v_y(y)$ and we extract $y_{\text{max}}$ for which $v_y$ is maximal, see Fig.~\ref{fitting_curve}-(b) which gives a typical example. In the same figure, $v_y(x,y=y_{\text{max}})$ is also plotted (red disks).

\begin{figure}[H]
\centering
\subfigure[]{\includegraphics[scale=0.33]{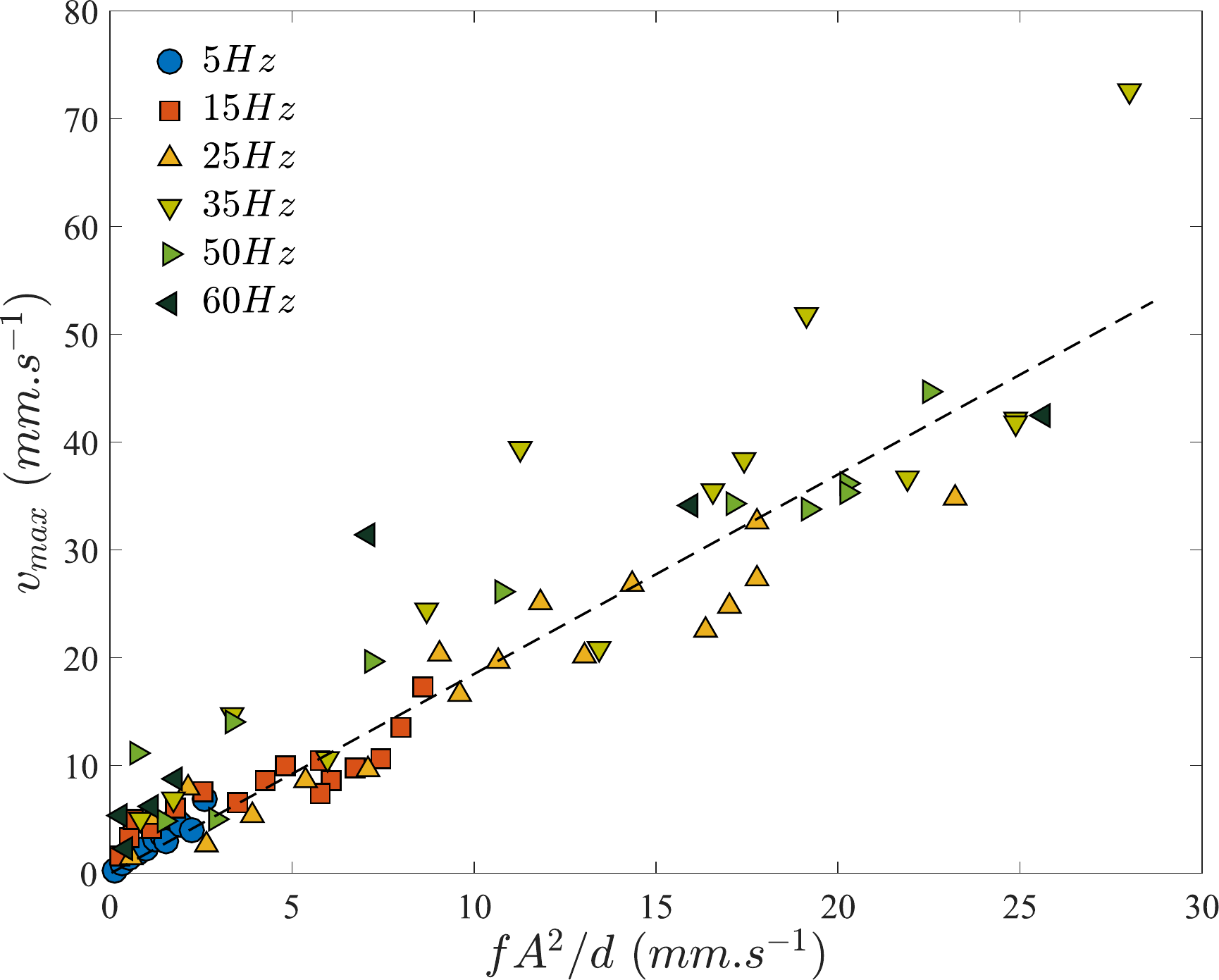}}
\subfigure[]{\includegraphics[scale=0.27]{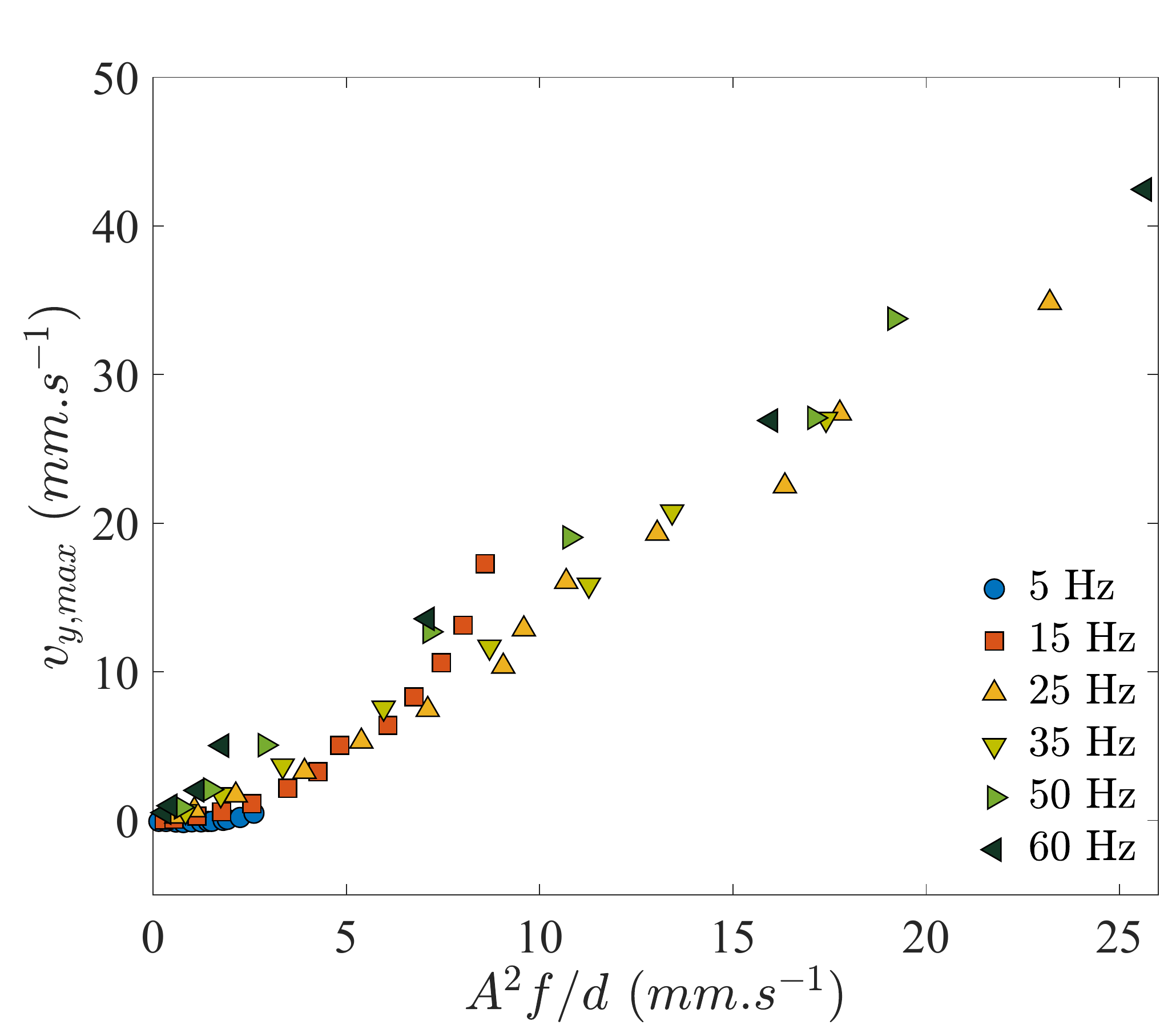}}
\subfigure[]{\includegraphics[scale=0.22]{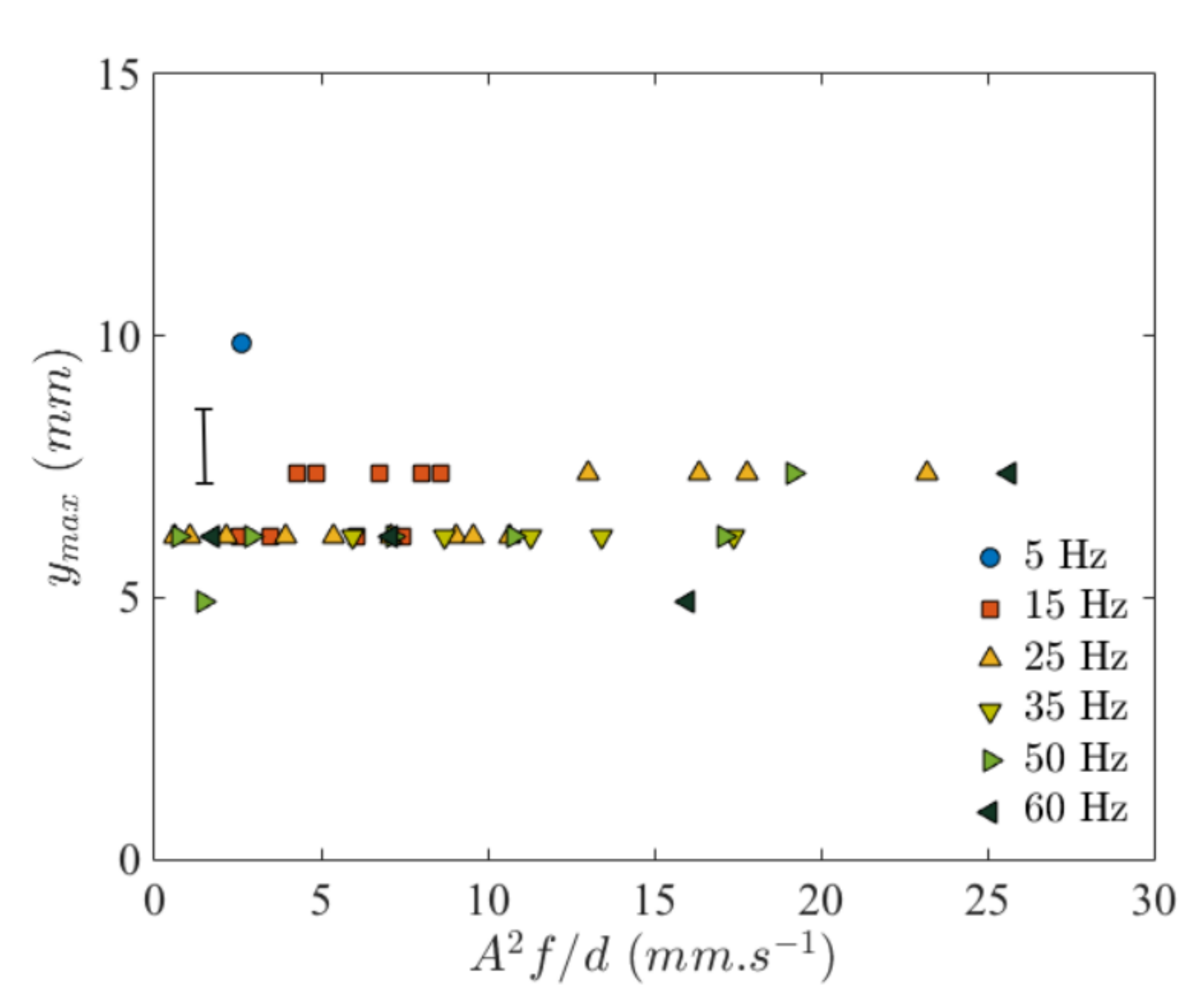}}
\subfigure[]{\includegraphics[scale=0.33]{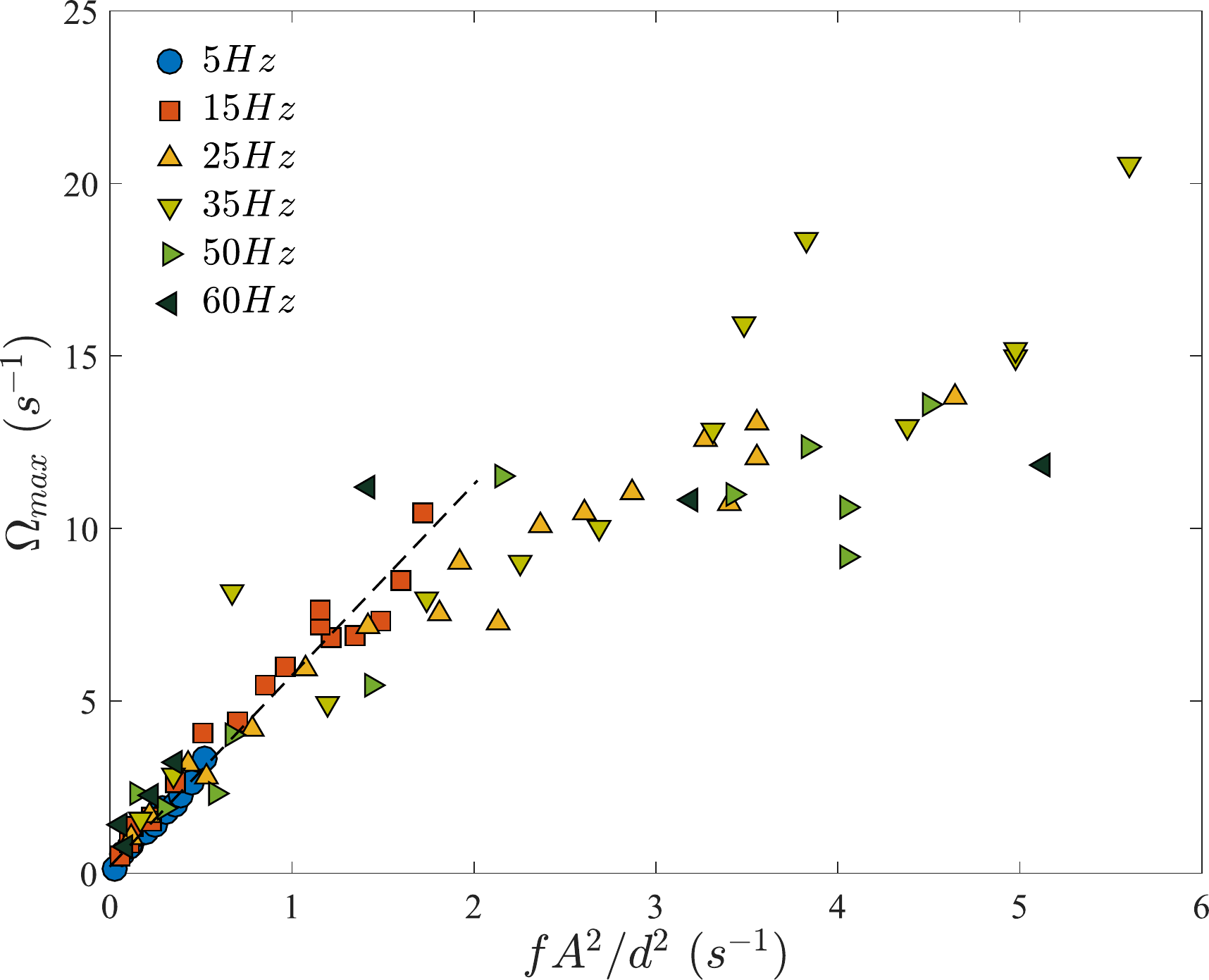}}
\caption{(a) Maximum streaming velocity versus $A^2f/d$ for different frequencies. (b) Maximum streaming velocity $v_{\text{max}}$ along the vibration $y$ axis, versus $A^2f/d$. (c) Location $y_{\text{max}}$ of the maximal velocity on the $y$ axis, with error bar. (d) Maximal vorticity versus $A^2f/d^2$.}
\label{fig:max_vw}
\end{figure}

The maximal velocity $v_{\text{max}}$, i.e. the maximal value of the norm of the velocity vector $ \Vert \vec{v_2} \Vert_{\text{max}}$ is plotted in Fig.~\ref{fig:max_vw}-(a) versus the characteristic velocity $A^2 f /d$, suggested from the scaling of eq.~(\ref{eq:scaling_v2}). Although some dispersion exists, these experiments are in fair agreement with eq.~(\ref{eq:scaling_v2}), even at relatively high forcing.

Figure \ref{fig:max_vw}-(b) shows the maximal velocity along the vertical axis (at $x$=0) $v_{y,\text{max}}$ versus characteristic velocity $A^2 f / d$, where the velocity has negligible component along $x$ by reasons of symmetry. These results are extracted from the same series of sequences as those of Figure \ref{fig:max_vw}-(a) - and show a much narrower dispersion. Remarkably again, the scaling law for $v_{y,\text{max}}$ holds well, even when $KC$ is slightly above 2, and the prefactor relating $v_{y,\text{max}}$ and $A^2 f / d$ is of the order of one. In Figure \ref{fig:max_vw}-(c), we plot the $y$ location for the maximal velocity $y_{\text{max}}$ versus $A^2 f / d$. This location remains roughly constant with the forcing amplitude and frequency - and is slightly larger than the object size. 

The maximal vorticity $\Omega_{\text{max}}$, determined within the outer boundary layer, is plotted versus $A^2 f /d^2$ in Figure \ref{fig:max_vw}-(d). Providing the forcing amplitude is not too large, these results are in agreement with the scaling of eqs.~(\ref{eq:scaling_v2}) and (\ref{eq:scaling_omega2}). They are also consistent with previous experiments on streaming generated with a vibrating beam in a 2D cell \cite{Costalonga15}. At high $A$, the velocity data show significant dispersion, and the maximal vorticity departs from the linear relationship $\Omega \sim f A^2/d^2$, that holds well at small enough $A$. 

To quantify the elongation of the secondary vortices, we naturally attempted to determine the position of the maximal vorticity (denoted as \\ $(x_m,y_m) \widehat{=} (r_m,\frac{\pi}{2} - \theta_m)$ in Cartesian or polar coordinates) or the position of the eyes of the vortices (which are actually that of the secondary flow streamlines), denoted as $(x_a,y_a) \widehat{=} (r_a,\frac{\pi}{2} - \theta_a)$ in Cartesian or polar coordinates. These positions are in general not equal: the maximas of vorticity are not located at the eye of the vortices, but close to the cylinder where velocity gradients are the strongest. It appears obvious in Figs.~\ref{fig:piv_vort}, and in general, $x_m < x_a$, $y_m < y_a$, $r_m < r_a$ and $\theta_m < \theta_a$.

\begin{figure}[H]
\centering
\subfigure[]{\includegraphics[scale=0.30]{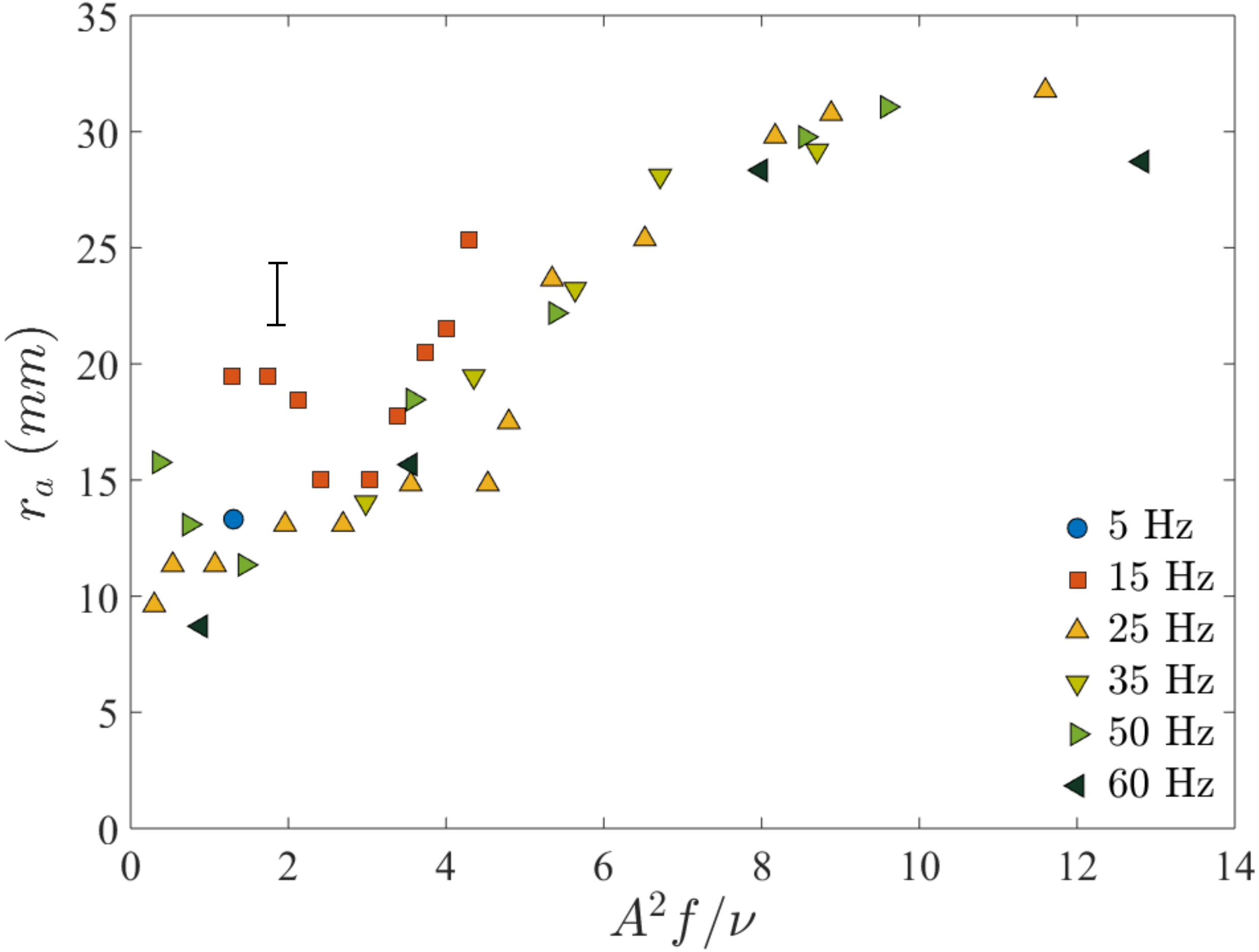}}
\subfigure[]{\includegraphics[scale=0.43]{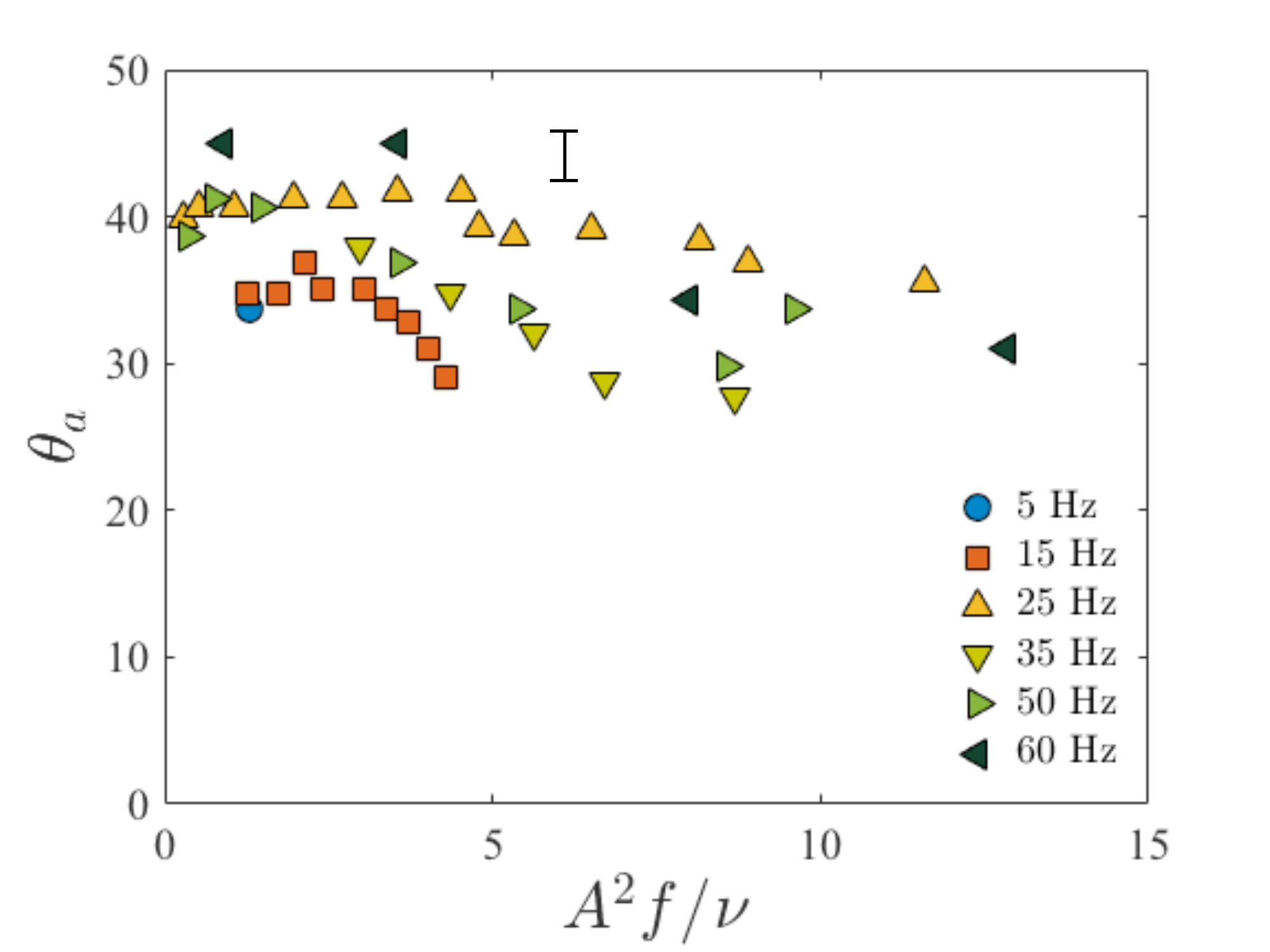}}
\caption{(a) Radial distance and (b) angular position of the center of secondary flow streamlines, versus $A^2 f / \nu$, at different frequencies. See definition in Fig.~\ref{fig:piv_vort}. Typical error bars are indicated.}
\label{fig:center_streamlines}
\end{figure}

Figures \ref{fig:center_streamlines} present the radial and angular position $(r_a,\theta_a)$ of the center of the streamlines, versus $Re_s (= A^2 f/\nu)$. Although the data is slightly dispersed, these results confirm that the flow streamlines stretch along the cylinder, along the direction of vibration, as the forcing gets stronger. Concerning the angle $\theta_a$, the data obtained at different frequencies do not collapse on a same curve. Overall, it slightly decreases from - roughly - 45$^{\circ}$ at low $Re_s$ down to 27$^{\circ}$ at larger $Re_s$. Although the position of the center of streamlines fairly restitutes the stretching effect, it does not constitute a quantitative enough measurement to reveal that the zones of intense vorticity get closer to the $y$ axis.

Hence, we attempted to determine the position of the maximal vorticity from the 2D vorticity maps, like those in Figs.~\ref{fig:piv_vort}, but due to the flatness of the velocity profiles, and the inherent noise of the experiments, the location of these maxima is subjected to large uncertainty. We propose an alternative way to quantify both the elongation and the narrowing of the vortices. We consider the spatial profile of the velocity decrease along the $y$ axis. This quantity is measurable with much better accuracy, as it results from a fit on more than fifteen data points in $v_y(y)$, as exemplified in Fig.~\ref{fitting_curve}-(b). Careful examination reveals that beyond the location of the maximal value $y_{\text{max}}$, the vertical velocity profile can be well fitted by the following empirical relationship:

\begin{equation}
v_y (x=0,y > y_{\text{max}}) = v_{y,\text{max}} \times \exp{\left(-\frac{y-y_{\text{max}}}{l_a} \right)}
\label{eq:fit_Vy}
\end{equation}

\noindent where the length $l_a$ quantifies the scale of the outer streaming flow: it is the distance along which the velocity decreases by a factor of $\frac{1}{e}$, from the location of the maximal velocity $y_{\text{max}}$, see Fig.~\ref{fitting_curve}-(b).

\begin{figure}[H]
\centering
\subfigure[]{\includegraphics[scale=0.4]{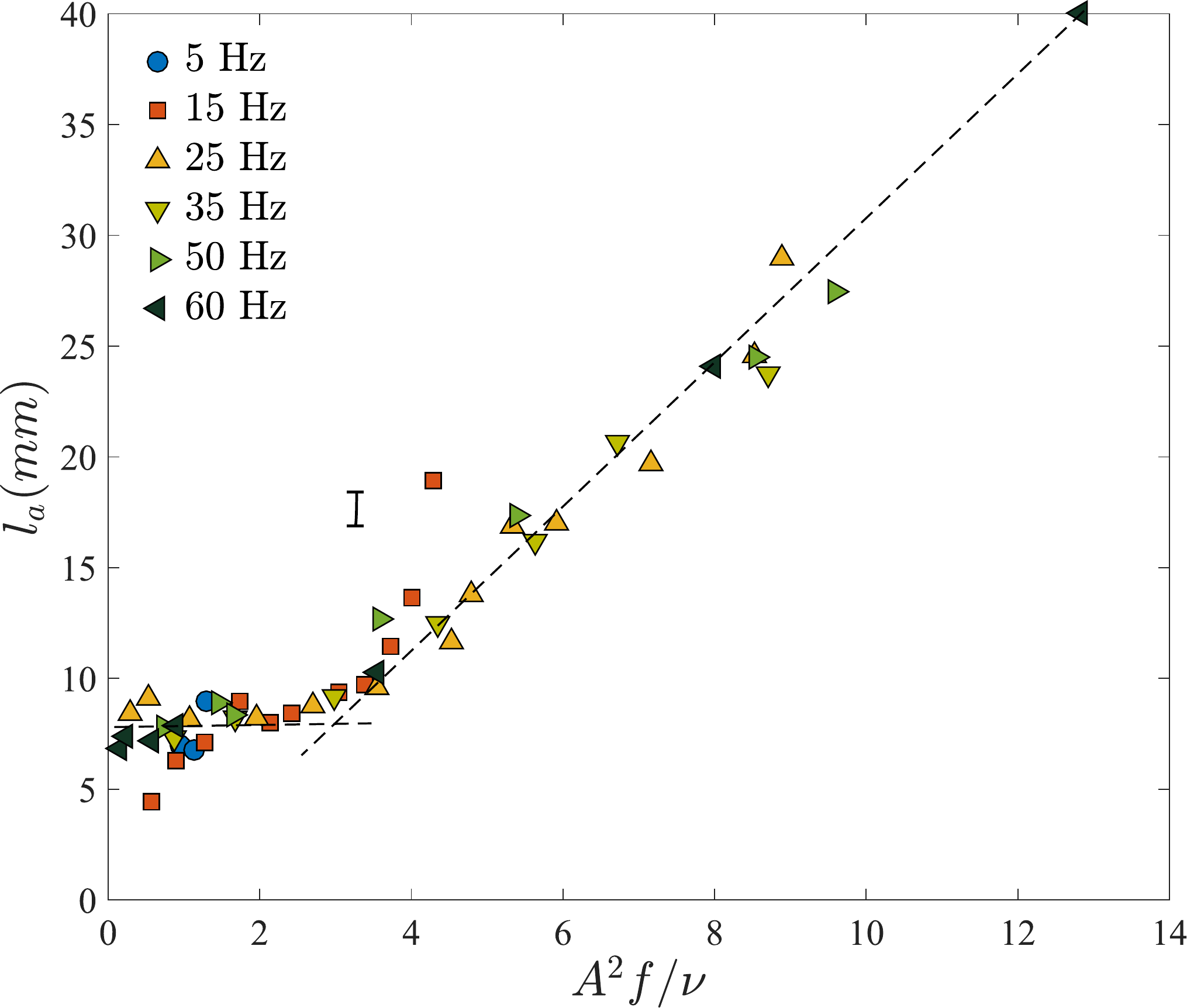}}
\subfigure[]{\includegraphics[scale=0.4]{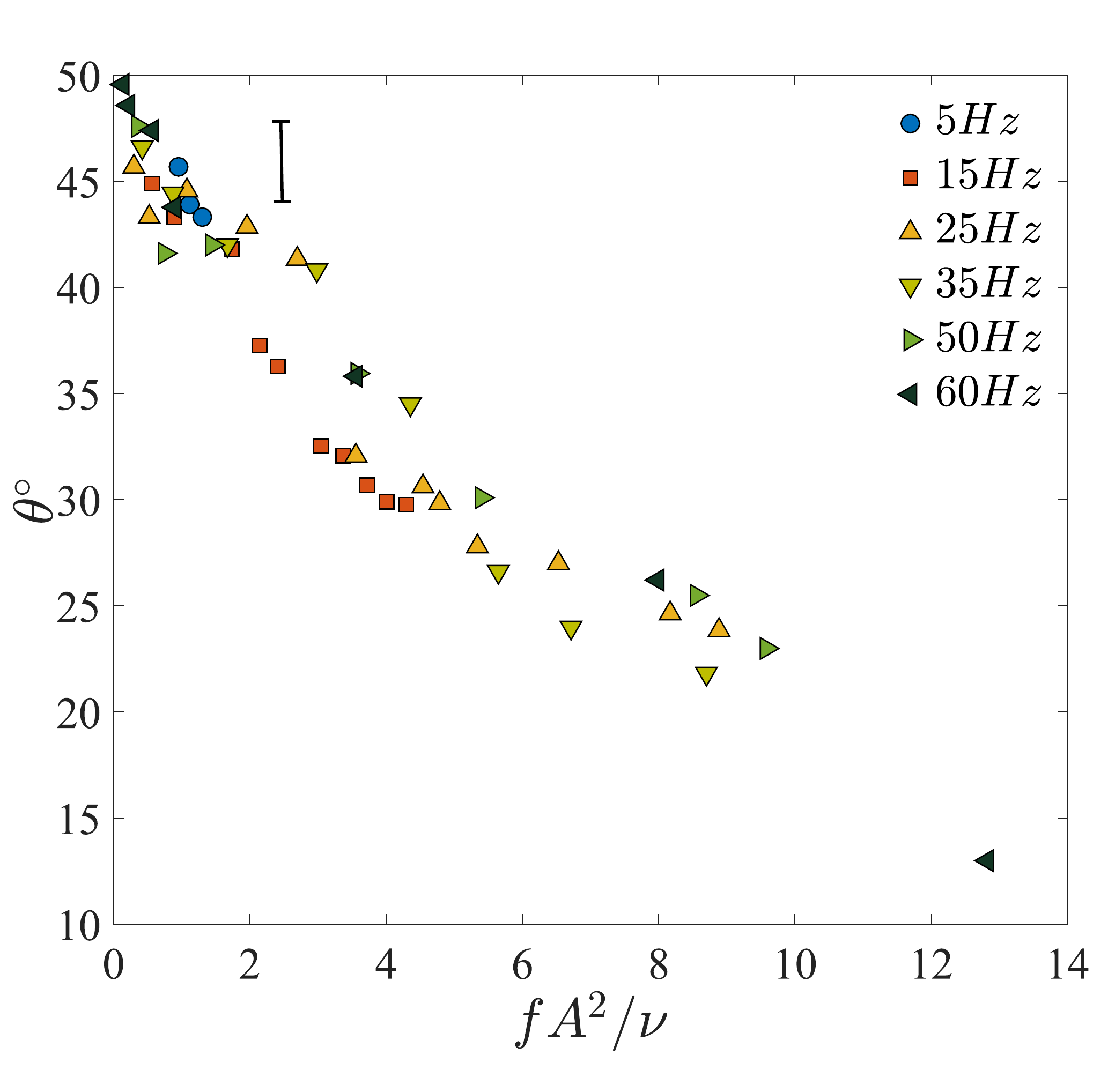}}
\caption{(a) Spatial scale of the streaming flow, quantified by the characteristic length $l_a$ of decrease of $v_y$ along the $y$ axis, versus $A^2 f / \nu$. (b) Angle $\theta$ of the direction for the farthest decrease of the vorticity with respect to the axis of vibration versus $A^2 f / \nu$ (see text for details). Typical error bars are indicated.}
\label{fig:flow_scale}
\end{figure}

Figure \ref{fig:flow_scale} shows measurements of this characteristic length $l_a$ extracted from PIV maps, versus the dimensionless quantity $Re_s = A^2 f /\nu$. These results reveal what was suggested by the direct observations of the PIV measurements and vorticity maps in Figs.~\ref{fig:piv_vort}. The typical length of the outer flow roughly equals 7 mm for Re$_s < 3$ and increases linearly: $l_a \sim A^2 f /\nu$, and reaches up to 8 times the length measured at low forcing.

Of course, the quantity $l_a$ is closely related to the location of the maximal vorticity, and to the size of outer vortices. At low enough $Re_s$, the size of vortices was found to be roughly equal to the size of the vibrating object \cite{Tatsuno77,Gondret_Petit,Costalonga15,Wang1968}, with a prefactor close to one, which shows $l_a$ to be a consistent quantity even at weak forcing. Hence, the stretching of streaming vortices along the direction of vibration $y$ seems related to the increase of the relative importance of inertia in the streaming flow. Still, their width in the $x$ direction remains roughly equal to $d$, which makes their aspect ratio significantly increasing with the forcing too.

We also evaluated the direction $\theta$ along which vorticity is the most intense. Indeed, from Figs.~\ref{fig:piv_vort}-(a,b), it is clear that the lobes of vorticity get closer to the (O,y) vibrations axis as the forcing increases. It is determined from vorticity maps like those in Fig.~\ref{fig:piv_vort}-(a,b), by measuring the angle corresponding to the farthest radial decrease of the vorticity. Although the fitting of data was not as neat as for the velocity $v_y$ with eq.~(\ref{eq:fit_Vy}), it was possible to unambiguously extract this angle $\theta$ with fair accuracy. 

Figure \ref{fig:flow_scale}-(b) represents this angle $\theta$ with respect to $(O,y)$, versus the streaming Reynolds number $Re_s = \frac{A^2 f}{\nu}$. While for $Re_s < $ 1, $\theta$ roughly equals 45$^{\circ}$, it significantly decreases for $Re_s >$ 1 and can reach a value down to 12$^{\circ}$, hence corresponding to a very narrow directional jet. At high forcing, the areas of large vorticity are much more aligned with $(O,y)$, and measurements for different frequencies from 5 Hz to 60 Hz collapse together well with the quantity $\frac{A^2 f}{\nu}$.

\section{DISCUSSION} 

Let us briefly summarize our results: the measurements of averaged and maximal vertical velocity are in agreement with classical scaling laws given by eq.~(\ref{eq:scaling_v2}) over the whole range of Re$_s$ investigated. Conversely, the maximal vorticity is well predicted by eq.~(\ref{eq:scaling_omega2}) only in a limited range of Re$_s$, typically below 3. For higher values of Re$_s$, $\Omega_{max}$ tends to saturate, see also Figs.~\ref{fig:evol_vort_25Hz}, which was not expected from classical theories. This is to be related to the elongation and narrowing of outer vortices.

From the previously presented results, the stretching of outer eddies suggests that the vorticity is transported further away from the cylinder, something already observed qualitatively for streaming flows of vibrating spheres \cite{Amin_Riley1990,Blackburn2002}. Still, the location of maximal velocity $y_{\text{max}}$ remains roughly fixed at a distance of the cylinder equal to $d$ - and, to the best of our measurement accuracy, so does the location to maximal vorticity ($r_m,\theta_m$). 

Stuart \cite{Stuart66} and Davidson and Riley \cite{Davidson_Riley72} predicted that the thickness of the outer boundary layer $D \sim d~(Re_s)^{-\frac{1}{2}}$, which can be simplified as $D \sim \frac{d}{A} \left( \frac{\nu}{\omega} \right)^{1/2} = \frac{d~\delta}{A}$. Furthermore, Stuart predicted that this outer layer was observed only if $Re_s $ is large enough, in practice equal or larger than a few units. Several experimental studies corroborated this latter prediction \cite{Davidson_Riley72,Bertelsen73,Tatsuno77}, although none of them quantitatively investigated the thickness of the outer layer. Davidson and Riley \cite{Davidson_Riley72} mentioned that the streaming flow narrows along the vibration axis at high Reynolds number. In most of these experiments, $A \simeq \delta$ and $A \ll d$. Let us also note that this elongation was already observed (though not quantitatively investigated), in a previous study carried out in a confined geometry \cite{Costalonga15}. Incidentally in this study, the saturation of $\Omega$ at high forcing (see Fig~\ref{fig:max_vw}-(d)) was related to the stretching of vortices, as the conservation of the circulation yields $\Omega_{\text{max}}\,d\,l_a \sim U_{str}\,d \sim A^2f$. Since $\Omega_{\text{max}}$ saturates at high forcing, $l_a$ must follow a $A^2 f$ increase to fulfil this conservation.


First, let us remind that the elongation starts to become significant when $Re_s = \frac{A^2 f}{\nu} = \left(\frac{A}{\delta}\right)^2 > 3$, hence when $\left(\frac{A}{\delta}\right) > 2$. This is precisely when the amplitude of oscillations start to become larger than the thickness of the oscillating boundary layer. In this situation, the vibrating object periodically pierces through the boundary layer, which leads to a significant time-dependent component of velocity out of the boundary layer. Hence, it is highly plausible that in this situation, a time-dependent component of the velocity exists outside the boundary layer of thickness $\delta$. This was confirmed here by our PIV measurements of the unsteady velocity fields at different phases of the oscillation. Still, our experiments also showed that there is no significant unsteady vorticity outside the boundary layer. This situation was already emphasized by Wang \cite{Wang1968}, who adopted the frame of decomposing the flow into an unsteady and a steady components, which can \textit{a priori} be of the same order. This frame is more adapted to the present study than the classical decomposition utilized under weak forcing amplitude, presented in the introduction. 

Following Wang's study \cite{Wang1968}, let us then decompose the velocity $\vec{v}$ and vorticity $\Omega$ fields as: 

\begin{eqnarray}
\vec{v} & = &\vec{v}_u + \vec{v}_s \\
\vec{\Omega} & = &\vec{\Omega}_u + \vec{\Omega}_s 
\label{eq:decomp}
\end{eqnarray}

\noindent where the subscripts u and s respectively stand for unsteady and steady components. Wang's final expression (see the detailed calculation in Appendix A) relates the stationary velocity $\vec{v_s}$ and vorticity $\vec{\Omega_s}$ with those of their unsteady counterparts, with dimensionless quantities:

\begin{equation}
- \left(\vec{\nabla} \times \vec{v}_u \times \vec{\Omega}_u \right)_s - \epsilon^2  \vec{\nabla} \times \vec{v}_s \times  \vec{\Omega}_s  =  \frac{\epsilon}{Re_1} \Delta \vec{\Omega}_s
\label{eq:wang}
\end{equation}

\noindent where $\epsilon = A/d$. The generation of vorticity usually comes from the first term, and the second one is neglected in most studies where $\epsilon \ll$1. While the first term corresponds to the generation of steady vorticity via the unsteady flow inside the viscous boundary layer, the second one corresponds to the convective transport of steady vorticity by the steady streaming flow itself. 

As stated above, the effect of this convective transport can be observed in Figure 3 of Tatsuno and Bearman's paper \cite{Tatsuno90}. A recent numerical study also seems to have reproduced this effect \cite{Nuriev18}. From the trends given by our experiments, we can assume that this mechanism can explain at least partially the observed vortex stretching for two main reasons:

- no significant unsteady vorticity could be observed outside the boundary layer (see Figs.~\ref{fig:unst_vort}), which keeps the first term of eq.~(\ref{eq:wang}) negligible outside the boundary layer and makes this term unlikely to be a source of vorticity in the outer domain $r > \delta$, even along the vibration axis. To check this point even further, we computed the term $\left(\vec{\nabla} \times \vec{v}_u \times \vec{\Omega}_u \right)$ for all phases, based on the unsteady velocity and vorticity fields already computed. The corresponding movie presented in Supplementary Information confirms that this term is negligible out of the boundary layer.

- the scaling of the length of stretching $l_a$ is very much related to that of the maximal streaming velocity $v_{y,\text{max}}$ along ($Oy$), see Figs.~\ref{fig:max_vw}-(b) and \ref{fig:flow_scale}-(a).

The decrease of the angle intersecting the middle of the vorticity lobes $\theta$ with increasing $Re_s$, could also be understood as that the strongly directional streaming flow would tend to narrow the lobes along the vibration axis. This distortion of the vortices was emphasized in \cite{Milton_Andres53} for the inner streaming vortices. The authors were attributing the distortion to fourth order terms (i.e. $\sim \epsilon^4$), generally neglected for low-amplitude forcing \cite{Holtsmark54}, which are contained here in the term $\left(\epsilon^2  \vec{\nabla} \times \vec{v}_s \times  \vec{\Omega}_s\right)$. 

\begin{figure}[H]
\centering
\includegraphics[scale=0.4]{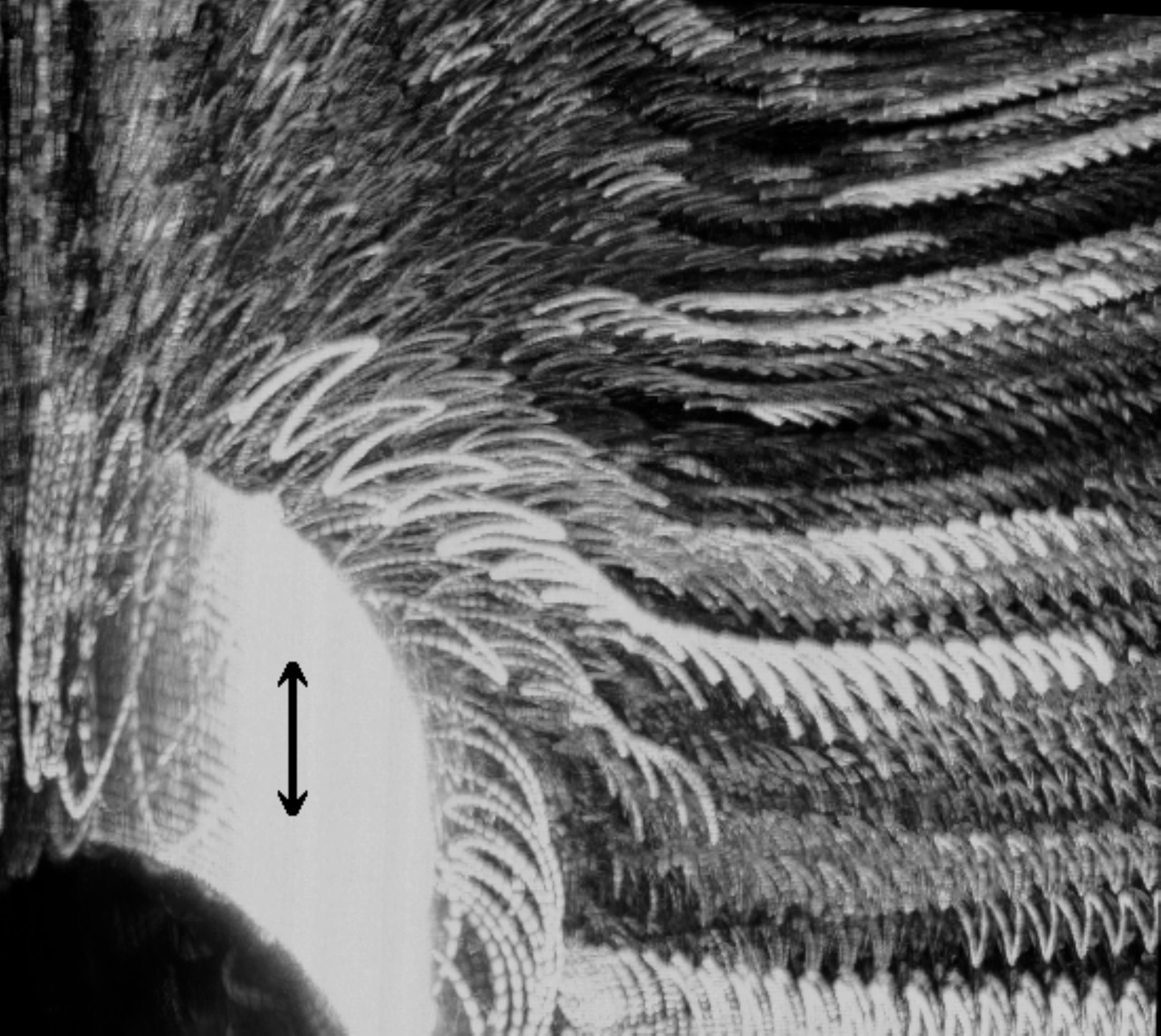}
\caption{Extraction of the unsteady flow trajectories during several consecutive periods, for same conditions as Figs. 5 to 7, which emphasizes that the unsteady and steady velocity can be comparable in magnitude. The arrow shows the direction of vibration.}
\label{fig:traj_unsteady}
\end{figure}

Finally, let us show another qualitative aspect of this convection of vorticity by extracting the trajectories of fluid particles from sequences at higher frames per second (here 2000 fps) during several forcing periods. A typical result is shown in Fig.~\ref{fig:traj_unsteady} obtained under the same conditions as in previous figures 5 to 7. The zig-zag like trajectories show that the unsteady and steady components of velocity can be comparable in magnitude, as assumed in the beginning of the discussion.

\section{CONCLUSION}

In conclusion, we investigated the steady streaming flow induced by vibrations of a cylinder immersed in a fluid, under conditions of high Reynolds number (still remaining in the $\mathcal{A}$* and $\mathcal{A}$ regimes defined by Tatsuno and Bearman \cite{Tatsuno90}) where the outer vortices thicken and stretch along the vibration axis. This phenomenon was noticed in previous experiments \cite{Davidson_Riley72,Tatsuno90}, but not quantitatively investigated nor explained. We reached regimes of large streaming Reynolds number $Re_s \gg$ 1, for which not only the typical flow velocity reaches up to a few cm/s, but also the outer vortices generated around the cylinder significantly stretch along the vibration axis up to 8 times the size they have at $Re_s <$1. The elongation comes together with a more directional secondary flow, as the angle of lobes of vorticity gets narrower along the axis of vibration. As the streaming velocity can be of the same order as the maximal unsteady one, we attribute the stretching to the convection of secondary vorticity by the streaming flow itself, an effect which is usually neglected within the frame of weak forcing.

The results shown in this study can be envisioned in a larger scope, including in applications of mixing and homogenization of fluids, which requires the generation of flow at a large distance from the source. Very recent studies focused on situations of a more complex geometry \cite{Coenen16}, microfluidics \cite{Nama14,Ovchinnikov14,Vishwanathan19} or nonlinear interactions between streaming and acoustic waves in acoustic streaming \cite{Daru17}. Depending on the size ratio between the mechanical actuator (or the acoustic wavelength) and the vessel, operating in conditions of high-Reynolds number flows can generate a flow whose spatial extension can be as large as the vessel or channel size. Hence, the resulting outer streaming flow can cover the whole fluid domain, and as a counterpart be also influenced and mediated by the walls via the no-slip condition, precisely as the inner streaming flow in viscous fluids \cite{Tatsuno77}.

\section*{APPENDIX A: DETAILS OF THE THEORY OF WANG \cite{Wang1968}}

Starting from the equation of transport of vorticity:

\begin{equation}
\frac{\partial \vec{\Omega}}{\partial t} - \vec{\nabla} \times \vec{v} \times  \vec{\Omega} = \nu \Delta \vec{\Omega}
\label{eq:cons_vort}
\end{equation}

This yields two equations for both steady and unsteady components: 

\begin{eqnarray}
\label{eq:vort_u}
\frac{\partial \vec{\Omega_u}}{\partial t} - \vec{\nabla} \times \vec{u}_u \times \vec{\Omega}_s -  \vec{\nabla} \times \vec{u}_s \times \vec{\Omega}_u - 
\left(\vec{\nabla} \times \vec{u}_u \times \vec{\Omega}_u \right)_u  =  \nu \Delta \vec{\Omega}_u \\
\label{eq:vort_s}
- \left(\vec{\nabla} \times \vec{u}_u \times \vec{\Omega}_u \right)_s - \vec{\nabla} \times \vec{u}_s \times \vec{\Omega}_s  =  \nu \Delta \vec{\Omega}_s
\end{eqnarray}

Let us mention that the first term in eq. (\ref{eq:vort_s}) comes from the steady component of the Reynolds stress generated by the first-order flow inside the time-periodic boundary layer. It is non zero, as the cross-product of two time-periodic terms can produce a time independent flow, i.e. with non-zero time-average.

Continuing in Wang's approach, we make dimensionless the above equations, by normalizing the following way (primed quantities denoting the dimensionless quantities): time by $1/\omega$, length by $d$, unsteady velocity by $V_0 = A \omega$ and steady velocity by $\gamma V_0$, $\gamma$ being a dimensionless ratio, unknown \textit{a priori}. It yields:

\begin{eqnarray}
\label{eq:vort_u_adim}
\bigg[ \frac{V_0 \omega}{d}\bigg] \frac{\partial \vec{\Omega'_u}}{\partial t} - \bigg[ \frac{\gamma V_0^2}{d^2} \bigg] \bigg(\vec{\nabla} \times \vec{u'}_u \times \vec{\Omega'}_s +  \vec{\nabla} \times\vec{u'}_s \times  \vec{\Omega'}_u \bigg) \nonumber \\
- \bigg[\frac{V_0^2}{d^2}\bigg]  \left(\vec{\nabla} \times \vec{u'}_u \times \vec{\Omega'}_u \right)_u  =   \bigg[ \frac{V_0}{d^3} \nu \bigg] \Delta \vec{\Omega'}_u  \, ,\\
\label{eq:vort_s_adim}
- \bigg[\frac{V_0^2}{d^2}\bigg] \left(\vec{\nabla} \times \vec{u'}_u \times \vec{\Omega'}_u \right)_s - \bigg[\frac{\gamma^2 V_0^2}{d^2}\bigg] \left(\vec{\nabla} \times \vec{u'}_s \times  \vec{\Omega'}_s \right)  =  \bigg[ \frac{\gamma V_0}{d^3} \nu\bigg] \Delta \vec{\Omega'}_s 
\end{eqnarray}

\noindent where the magnitude of the different terms are emphasized with brackets. Dropping the primes on the dimensionless quantities, and dividing the first above equation by $\left(\frac{V_0 \omega}{d} \right)$ and the second above equation by $\left(\frac{V_0^2}{d^2} \right)$, it yields:

\begin{eqnarray}
\label{eq:vort_u_ReS}
\frac{\partial \vec{\Omega_u}}{\partial t} - \frac{\gamma}{S}\bigg(\vec{\nabla} \times \vec{u}_u \times \vec{\Omega}_s + \vec{\nabla} \times \vec{u}_s \times  \vec{\Omega}_u \bigg) \nonumber \\
- \frac{1}{S} \left(\vec{\nabla} \times \vec{u}_u \times \vec{\Omega}_u \right)_u  = \frac{1}{Re_1 S}\Delta \vec{\Omega}_u \\
\label{eq:vort_s_ReS}
- \left(\vec{\nabla} \times \vec{u}_u \times  \vec{\Omega}_u \right)_s - \gamma^2  \vec{\nabla} \times \vec{u}_s \times  \vec{\Omega}_s  =  \frac{\gamma}{Re_1} \Delta \vec{\Omega}_s
\end{eqnarray}

\noindent defining $S = \frac{d \omega}{V_0} = \frac{1}{\epsilon}$ as the Strouhal number. As the product $Re_1 S = M^2 \gg$ 1, the equation for unsteady terms (\ref{eq:vort_u_ReS}) shows that, by dropping non-linear terms, a boundary layer of $O (Re_1S)^{-1/2}$ outside which the unsteady vorticity decays. In order to balance this forcing term in the steady equation (\ref{eq:vort_s_ReS}), the diffusive term must be of the order $1/(Re_1 S)$, which implies:  $\gamma =  \frac{1}{S} = \epsilon$, from which eq. (\ref{eq:wang}) is obtained.

\end{document}